\documentclass[iop]{emulateapj}

\usepackage{comment}
\usepackage{amsmath}
\usepackage{graphicx,color}
\usepackage{scrextend}
\usepackage{hyperref}
\usepackage{indentfirst}
\usepackage{verbatim}
\usepackage{morefloats}
\usepackage{natbib}
\usepackage{epstopdf}
\usepackage{appendix}
\usepackage{xcolor}
\definecolor{dgreen}{RGB}{26,148,49}

\def\beq{\begin{equation}}
\def\eeq{\end{equation}}

\def\bea{\begin{eqnarray}}
\def\eea{\end{eqnarray}}

\def\pa{\partial}
\def\OGD{\mathcal{O}^{\rm GD}}
\def\OIM{\mathcal{O}^{\rm IM}}
\def\Oopt{\mathcal{O}^{\rm opt}}
\def\OGDhat{\hat{\mathcal{O}}^{\rm GD}}
\def\OIMhat{\hat{\mathcal{O}}^{\rm IM}}

\shorttitle{Optimally mapping Large-Scale Structures with luminous sources}
\shortauthors{Cheng et al.}

\begin{document}

\title{Optimally mapping Large-Scale Structures with luminous sources}

\author{Yun-Ting Cheng$^1$, Roland de Putter$^{1}$, Tzu-Ching Chang$^{1,2,3}$,  Olivier Dor{\'e} $^{1,2}$}
\address{$^1$California Institute of Technology, 1200 E. California Blvd., Pasadena, CA 91125, U.S.A.}
\address{$^2$Jet Propulsion Laboratory, California Institute of Technology, 4800 Oak Grove Drive, Pasadena, CA 91109, U.S.A.}
\address{$^3$Institute of Astronomy and Astrophysics, Academia Sinica, 1 Roosevelt Rd, Section 4, Taipei, 10617, Taiwan}

\begin{abstract}
Intensity mapping has emerged as a promising tool to probe the three-dimensional structure of the universe. The traditional approach of galaxy redshift surveys is based on individual galaxy detection, typically performed by thresholding and digitizing large-scale intensity maps. By contrast, intensity mapping uses the integrated emission from all sources in a 3D pixel (or voxel) as an analog tracer of large-scale structure. In this work, we develop a formalism to quantify the performance of both approaches when measuring large-scale structures. We compute the Fisher information of an arbitrary observable, derive the optimal estimator, and study its performance as a function of source luminosity function, survey resolution, instrument sensitivity, and other survey parameters. We identify regimes where each approach is advantageous and discuss optimal strategies for different scenarios. To determine the best strategy for any given survey, we develop a metric that is easy to compute from the source luminosity function and the survey sensitivity, and we demonstrate the application with several planned intensity mapping surveys.
\footnote{\textcircled{c} 2018. All rights reserved.}

\end{abstract}


\section{Introduction}
Studying the large-scale structure (LSS) of the universe is a major focus in cosmology. The initial conditions of the LSS have been well characterized from the cosmic microwave background (CMB) measurements \citep[e.g.,][]{2016A&A...594A..13P,2018arXiv180706209P}, and powerful constraints on the cosmological parameters have been inferred from its measurement. Nevertheless, to map LSS at late time is an essential cosmological probe, in particular regarding the properties of dark matter and dark energy. 
By detecting a large number of individual galaxies as tracers of the underlying density field, one can map out the large-scale matter distribution and infer powerful cosmological constraints from its power spectrum, for example. This galaxy detection (GD) approach has been successfully demonstrated by several major observational programs such as 2dF \citep{2003astro.ph..6581C}, 6dF \citep{2009MNRAS.399..683J}, WiggleZ \citep{2012PhRvD..86j3518P}, VIMOS \citep{2014A&A...566A.108G}, SDSS \citep{2000AJ....120.1579Y}, and BOSS \citep{2013AJ....145...10D}. Upcoming galaxy surveys are expected to provide further unparalleled cosmological insights, e.g., eBOSS \citep{2016AJ....151...44D}, DESI \citep{2016arXiv161100036D}, PFS \citep{2014PASJ...66R...1T}, Euclid \citep{2011arXiv1110.3193L}, LSST \citep{2009arXiv0912.0201L}, WFIRST \citep{2015arXiv150303757S}, and SPHEREx \citep{2014arXiv1412.4872D}.

At higher redshift, GD becomes difficult, as galaxies at earlier times are on average fainter, and the increased distance reduces the observed flux. As a result, to detect a given number of galaxies at high redshift requires a longer integration time. This has in part motivated the development of intensity mapping (IM) as an alternative technique to probe LSS. Without thresholding to identify individual sources, IM traces the underlying density field using the integrated light emission from all the sources, including unresolved faint galaxies (see \citet{2017arXiv170909066K} for a recent review). In addition, line intensity mapping probes the three-dimensional structure by mapping the emission of a particular spectral line and uses the frequency-redshift relation to infer the matter distribution along the line of sight. The 21cm hyperfine emission from neutral hydrogen  \citep{1990MNRAS.247..510S,1997ApJ...475..429M,2008MNRAS.383..606W,2008PhRvL.100i1303C}, the CO rotational lines \citep{2008A&A...489..489R,2010JCAP...11..016V,2011ApJ...730L..30C,2011ApJ...741...70L,
2011ApJ...728L..46G,2014MNRAS.443.3506B, 2013ApJ...768...15P, 2015JCAP...11..028M,2015ApJ...814..140K,2016MNRAS.457L.127B,2016ApJ...817..169L,2016ApJ...830...34K,2017MNRAS.464.1948F,2017MNRAS.468..741B,2019ApJ...872..186C}, the [CII] 157.7 $\mu$m fine-structure line \citep{2012ApJ...745...49G,2014ApJ...793..116U,2015ApJ...806..209S,2015MNRAS.450.3829Y,2017MNRAS.464.1948F}, and the Lyman-$\alpha$ emission line \citep{2013ApJ...763..132S,2014ApJ...785...72G,2014ApJ...786..111P,
2016MNRAS.455..725C,2016MNRAS.457.3541C,2017MNRAS.464.1948F,2018MNRAS.481.1320C} are amongst the most studied lines in the IM regime.

Although the measurement can be challenged by the presence of continuum foregrounds \citep[e.g.,][]{2006PhR...433..181F,2006ApJ...648..767M,2009ApJ...695..183B,
2012MNRAS.419.3491L,2012ApJ...756..165P,2012MNRAS.423.2518C,2015ApJ...815...51S}, and line interlopers \citep{2016ApJ...825..143L,2016ApJ...832..165C}, it is still anticipated that line intensity mapping can provide an efficient path to access the faint, high-redshift Universe owing to its relatively low requirement on spatial resolution and sensitivity, which enables the use of small apertures to  efficiently scan a large comoving volume. 

The first measurement of IM signals from LSS used the 21 cm line. The detection was made in cross-correlation with spectroscopic galaxy catalog \citep{2010Natur.466..463C,2013ApJ...763L..20M,2018MNRAS.476.3382A}, and  auto-power spectrum constraints have been reported in \citet{2013MNRAS.434L..46S}. \citet{2013ApJ...768...15P} made the first attempt at measuring CO IM signal in cross-correlation but detected no signal. The COPSS II experiment measured the CO auto-power spectrum at z$\sim$3 \citep{2016ApJ...830...34K}. A tentative [CII] measurement has been made by \citet{2018MNRAS.478.1911P} in cross-correlation. While \citet{2016MNRAS.457.3541C} reported a first detection of Ly$\alpha$ emission in the IM regime by cross-correlating SDSS spectra with a quasar sample, a new analysis in \citet{2018MNRAS.481.1320C} using cross-correlation with both quasars and Ly$\alpha$ forest showed no detection of diffuse Ly$\alpha$ emission.

Formally, the main difference between GD and IM resides in the `weighting' of the observed data. In GD, the universe is digitized into a binary map where detected galaxies have a weight of one, and zero elsewhere. This is essentially giving a uniform weight to all the detected sources, regardless of their flux. On the contrary, IM is a linear mapping between the universe and the data, weighted by the observed intensity. These two different options are suitable for gleaning more information from the data in two extreme regimes: GD is ideal in the high spatial/spectral resolution and deep integration limit, where detected sources are less susceptible to the effects of noise and  confusion; IM is ideal if the individual voxel intensity is composed of highly confused sources with a non-negligible noise component.

In this work, we formally explore this dichotomy by introducing an ``observable,'' $\hat{\mathcal{O}}$, and quantify the information that can be extracted using this observable for a given survey using the Fisher information formalism. The GD and IM approaches represent two special cases of $\hat{\mathcal{O}}$. We define an ``optimal observable''  that optimizes the information extraction, not necessarily limited to the usual GD or IM approaches. We further develop a simple diagnostic to evaluate the two strategies (e.g. GD or IM) for a survey. We then apply this method to optimize survey design for future experiments and, as an example, optimize the pixelization of intensity maps considering two different noise levels. 

This paper is organized as follows. We first introduce our mathematical formalism in Sec.~\ref{S:formalism} before discussing two toy models within this formalism in Sec.~\ref{S:toymodel}. Scenarios with a more realistic model based on the Schechter luminosity function model are presented in Sec.~\ref{S:SchLF}. We then follow with two applications of our framework: we determine the optimal observable for several future planned surveys in Sec.~\ref{S:surveys}, and we optimize the survey pixel size in Sec.~\ref{S:OptPix}. The conclusions are given in Sec.~\ref{S:conclusion}.

\section{Formalism}\label{S:formalism}

A major goal of large-scale galaxy or intensity mapping surveys is to use emission from luminous sources to trace the underlying density field. In particular, we are interested in the {\it matter} overdensity field $\delta({\bf x}) \equiv (\rho({\bf x})-\bar{\rho})/\bar{\rho}$, where $\rho({\bf x})$ is the local matter density and $\bar{\rho}$ its  mean on large scales, from which cosmological information can be extracted (e.g., using the power spectrum statistics).
We can use luminous sources to learn about $\delta$ because, on large scales,
the overdensity of a sample of galaxies is a linearly biased tracer of the underlying matter density. In other words, neglecting stochastic noise, on large scales we have
\begin{equation}
\delta_g({\bf x}) \equiv (n_g({\bf x}) - \bar{n}_g)/\bar{n}_g = b \, \delta({\bf x}),
\end{equation}
where $n_g({\bf x})$ is the number density of a sample of galaxies at position ${\bf x}$, $\bar{n}_g$ its global mean, and $b$ the galaxy bias.

However, we do not observe $n_g$ directly, but the light emitted by galaxies. For a wide range of survey scenarios, 
we simply have access to the observed fluxes $L$ in many pixels or voxels, typically small in comparison to the large-scale overdensity modes of interest. These fluxes may include contributions from multiple luminous sources. The question we will tackle is how to optimally  extract $\delta$ from this ``data cube'' composed of these small pixels/voxels.

The terms 'pixel' and 'voxel' above respectively refer to a spatial 2D resolution element or a spatial-spectral 3D resolution volume element. Voxels are the data element in 3D line intensity mapping. A voxel volume can be written as $V_{\rm vox}\propto \Omega_{\rm pix}\,\Delta\nu$, where $\Omega_{\rm pix}$ is the solid angle of the angular size of a voxel and $\Delta\nu$ is the wavelength or frequency width. $\Omega_{\rm pix}$ and $\Delta\nu$ are usually chosen to be of the order of the survey point spread function (PSF; or beam size) and spectral resolution, respectively. However, the analysis in this work is not necessarily limited to the original voxel configuration of a given survey, as voxel size can always be increased by rebinning.

For simplicity, we assume that every source in the surveys fills in at most a single voxel, i.e., all the flux from a given source is measured in only one voxel, so that the correlation between voxels only arises from the underlying cosmological signal, i.e., source clustering. 
This assumption requires that the voxel size to be at least a few times larger than the PSF (beam) size and the size of the sources themselves. Likewise, in the spectral dimension, we require the voxel size to be larger than a few times of the spectral resolution and the target line width. Alternatively, the analysis in this work also applies to 2D imaging of a single frequency band. In this case, a 3D voxel reduces to a 2D pixel, and we also require the pixel size to be a few times larger than the beam size.

\subsection{Observables}\label{S:observables}

To extract information about the underlying cosmological matter overdensity, we consider a general ``observable function,'' $\mathcal{O}(L)$, serving as a weight function turning the observed map of voxel fluxes\footnote{The unit of flux $L$ in each voxel is power per area, in [W m$^{-2}$] (or [photons s$^{-1}$ m$^{-2}$]). $L$ is an ``extensive'' quantity under this definition, i.e. its value is scaled with the voxel size. Furthermore, later in the paper we will directly compare $L$ with the intrinsic luminosity (in units of $W$ or $L_\odot$) of the sources $\ell$. In this case, we implicitly assume that $\ell$ has been converted to the flux $\ell/4\pi D_L^2$ such that the two quantities are in the same units.
} $\hat{L}$ into a transformed ``observable map'' with values $\hat{\mathcal{O}} \equiv \mathcal{O}(\hat{L})$ in each voxel\footnote{Throughout the paper, we use the hat notation as a specific realization of the quantity. Thus, $L$ is a variable, while $\hat{L}$ refers to a specific realization of $L$. Likewise, $\mathcal{O}(L)$ refers to function $\mathcal{O}$ with variable $L$, and $\hat{\mathcal{O}}$ is the function value at $L=\hat{L}$.}. The power spectrum of this new $\mathcal{O}(\hat{L})$ map is then computed as a proxy for the underlying overdensity field matter density power spectrum.

As an alternative way of thinking about how the voxel map can be used to constrain the large-scale matter overdensity, we consider a region that is small compared to the matter overdensity long-wavelength modes of interest so that, in this region, the $\delta$ of the long-wavelength modes  is nearly uniform (i.e., it can be treated as a ``DC mode''). We can further assume the voxel scale to be much smaller than the scale of the long-wavelegth cosmological modes of interest, so we may choose our local region such that it still contains a large number of voxels. In this picture, the way the local overdensity $\delta$ is constrained is using the sum (or average) of the values of $\hat{\mathcal{O}}$ in the voxels in the local region.

In our context, the quantity of interest is the ``large-scale'' rather than ``total'' density field. In principle, each voxel traces the ``total'' underlying density field, $
\delta_{\rm tot}$, which is composed of both large- and small-scale fluctuations: $\delta_{\rm tot}  = \delta_L + \delta_S$. Here we only constrain $\delta_L$ through the average value of the observable $\hat{\mathcal{O}}$ over a large number of voxels living in approximately the same local $\delta_L$, i.e. we use an ensemble average of $\langle \mathcal{\hat{O}} \rangle$, not individual voxel measurement $\mathcal{\hat{O}}$, to trace the large-scale fluctuation $\delta_L$. Since $\delta_L$ does not refer to a specific scale of fluctuation, this argument applies to any modes that have a fluctuation scale much greater than the voxel size. We will thus write $\delta$ instead of $\delta_L$ from now on, but the readers should keep in mind that the $\delta$ we discuss in this work does not include small-scale fluctuations $\delta_S$.

GD and IM represent two special cases of such a mapping $\mathcal{O}(L)$. For GD, a voxel is labeled as a ``detection'' if it is brighter than a threshold luminosity $L_{\rm th}$ (say, 5 times the noise rms for a $5\sigma$ detection). A power spectrum can then be calculated with this ``digital map'' that consists of ones (detection) and zeros (nondetection) with a proper normalization. Therefore, $\mathcal{O}(L)$ in this case is a step function at $L_{\rm th}$,
\beq
\mathcal{O}^{{\rm GD}}(L) =
\begin{cases}
1  \quad \quad \text{if} \, L > L_{\rm th}\\
0 \quad \quad \text{if} \, L \leq L_{\rm th}.
\end{cases}
\eeq
On the contrary, IM directly calculates a power spectrum of the measured intensity (or luminosity) map, so the observable is a linear function of $L$ (the trivial, identity map), 
\beq
\mathcal{O}^{\rm IM}(L) = L.
\eeq

While the observed fluxes contain a wealth of additional information (for instance, on galaxy evolution and small-scale clustering), we focus our study on how to optimally extract the underlying cosmological matter overdensity $\delta$. Let's consider a fixed realization of the overdensity $\delta$ in some region containing many voxels. A given choice of observable $\mathcal{O}(L)$ leads to a noisy estimate of the local value of $\delta$, where the noise is due to the shot noise in the source population used as density tracers and to the instrumental noise. In practice, we will aim at minimizing the combined noise. Our final goal is to measure the large-scale power spectrum of the observable map $\hat{\mathcal{O}}$. Uncertainties in the power spectrum contain a cosmic variance component (signal), due to the variance in the underlying matter overdensity $\delta$, and a stochastic/shot-noise component, which is given by how well the observed fluxes from luminous tracers measure the underlying cosmological clustering. By minimizing the noise in the local determination of $\delta$, we minimize the stochastic noise power spectrum relative to the cosmic variance part of the power spectrum, which is the signal of interest.

We will quantify the maximum information content of $\delta$ by its Fisher information. We will show that there exists an ``optimal observable'' $\mathcal{O}^{\rm opt}$ such that this observed map contains the same amount of information as the Fisher information. The functional form of this optimal observable depends on the voxel luminosity probability density function (pdf) and we detail its derivation in Sec.~\ref{S:voxPDF} before describing in Sec.~\ref{S:FisherOpt} the Fisher information and optimal observable.

\subsection{Voxel Luminosity pdf}\label{S:voxPDF}

The voxel luminosity pdf $P(L,\delta)$ is defined as the probability of a voxel residing in an overdensity field $\delta$ with a luminosity between $\left[L, L+dL\right]$. This can be computed by the $P(D)$ analysis presented in \citet{2009JCAP...07..007L}. First, we define $P_k(L,\delta)$ to be the probability of the voxel with luminosity between $\left[L, L+dL\right]$ given that there are $k$ sources in that voxel. The $P(L,\delta)$ is the summation of all the $P_k(L,\delta)$  weighted by the probability of occurrence of each $k$. If the sources are uncorrelated, the weight function is a Poisson distribution, and thus
\begin{equation}\label{E:PL}
P(L,\delta)=\sum_{k=0}^{\infty} \frac{e^{-N(\delta)}\,N^k(\delta)}{k!}\,P_k(L,\delta),
\end{equation}
where $N(\delta)$ is the expectation value of the number of sources in a voxel with overdensity $\delta$. The clustering effects can be accounted for by modifying the Poisson term in Eq.~\ref{E:PL}, for example, the approaches presented in \citet{2017MNRAS.467.2996B}. For simplicity
in this work, we only adopt the Poisson distribution in the $P(L)$ function, and we leave the consideration of clustering to future work.

$N(\delta)$ and $P_k(L,\delta)$ can be derived for any given luminosity function\footnote{Throughout this paper, $L$ refers to the total luminosity in a voxel, and $\ell$ denotes the luminosity of a single source.} $\Phi(\ell,\delta)$ and voxel volume $V_{\rm vox}$,
\begin{align}
&N(\delta)=V_{\rm vox}\int\Phi(\ell,\delta)\,d\ell,\label{E:Ndef}\\
&P_0(L,\delta)=\delta^D(L),\\
&P_1(L,\delta)=\Phi(L,\delta)/\int \Phi(\ell,\delta)\,d\ell,\\
&P_k(L,\delta)=\int P_1(L',\delta)\,P_{k-1}(L-L',\delta)\,dL'.
\end{align}

The effect of instrumental noise can be easily included by convolving $P(L,\delta)$ with the noise pdf. In this work, we only consider Gaussian noise with a constant rms $\sigma_L$ that does not depend on the intrinsic luminosity, so the noisy $P(L,\delta,\sigma_L)$ is given by\footnote{To simplify the notation, we will drop the $\sigma_L$ notation in $P(L,\delta,\sigma_L)$ in the following paper unless it is helpful to clarify in certain situations.}, 
\begin{equation}\label{E:PLG}
\begin{split}
P&(L,\delta,\sigma_L)=P(L,\delta)\ast G(\sigma_L)\\
&\equiv \int dL'\,P(L',\delta)\,\frac{1}{\sqrt{2\pi}\,\sigma_L}\,e^{-(L-L')^2/2\sigma_L^2} .
\end{split}
\end{equation}

Throughout this paper we consider multiple values of $N\equiv N(\delta=0)$, the mean number of sources per voxel, given in Eq.~\ref{E:Ndef}. We note that variations in $N$ can be interpreted in two useful ways. First, a change in $N$ can represent a change in the number of objects for a fixed voxel size, i.e., a change in the amplitude of the luminosity function $\Phi(\ell)$ describing the source population. Alternatively, it is often instructive to consider a change in $N$ as a change in the voxel volume, $V_{\rm vox}$, for a fixed physical source population. This allows us to study information content vs.~voxel size. In the latter case, the noise per voxel, $\sigma_L$, may of course also vary as voxel size or $N$ is varied.

\subsection{Fisher Information}\label{S:FisherOpt}

Assuming that voxels are independent tracers of the large-scale density field $\delta$, the likelihood of the whole measurement is the product of the likelihood over all voxels $i$, $P(\hat{L}_i,\delta)$, (Eq.~\ref{E:PL}), 
\beq\label{E:likelihood}
\mathcal{L}(\{\hat{L}_i\};\delta)=\prod_{i} P(\hat{L}_i,\delta).
\eeq
The full Fisher information content on $\delta$ of this whole measurement is defined as \citep{2001MNRAS.328.1027T}
\beq
F^{\rm full}_{\delta\delta}=-\langle \partial^2_{\delta}\ln \mathcal{L}(\{\hat{L}_i\};\delta)^2 \rangle=\langle (\partial_{\delta}\ln \mathcal{L}(\{\hat{L}_i\};\delta))^2 \rangle,
\eeq
where $\langle f\rangle=\int dL\, P(L,\delta)\,f(L)$ is the expectation value of function $f$. The Cram{\'e}r-Rao inequality states that $\sigma_{\delta}^2\geq 1/F^{\rm full}_{\delta\delta}$, thus placing a lower bound on the variance of parameter $\delta$ that one can attain with the data \citep{1997ApJ...480...22T}. 
Using Eq.~\ref{E:likelihood}, we get 
\begin{equation}
F^{\rm full}_{\delta\delta}=\langle (\partial_{\delta}\ln [\prod_{i} P(\hat{L}_i,\delta)])^2 \rangle
=\sum_i\langle (\partial_{\delta}\ln P(\hat{L}_i,\delta))^2 \rangle,
\end{equation}
and thus $F_{\delta\delta}\equiv \langle (\partial_{\delta}\ln P(\hat{L}_i,\delta))^2 \rangle$ is the \textit{total Fisher information content per voxel}. Below we will quantify the Fisher information of this per-voxel basis.

In the context of this work, the parameter $\delta$ is estimated from the mean value of observable map $\hat{\mathcal{O}}$ over a large amount of voxel data. In this case, the Fisher information per voxel for this observable is \citep{2013MNRAS.434.2961C}
\begin{equation}\label{E:Fisher}
F^\mathcal{O}_{\delta\delta}=\frac{\left( \partial_{\delta} \left \langle \mathcal{\hat{O}}\right \rangle\right)^2}
{\left \langle \mathcal{\hat{O}}_{}^2 \right \rangle-\left \langle \mathcal{\hat{O}} \right \rangle^2}=\frac{\left( \partial_{\delta} \left \langle \mathcal{\hat{O}}\right \rangle\right)^2}
{\sigma^2({\hat{\mathcal{O}}})},
\end{equation}
where the denominator $\sigma^2({\hat{O}})$ is the variance in map $\hat{\mathcal{O}}$ per voxel and $\left \langle \cdot \right \rangle$ is the expectation value defined above. The condition $F^\mathcal{O}_{\delta\delta}\leq F_{\delta\delta}^{}$ holds, as the Fisher information extracted with any given observable cannot exceed the total Fisher information content. The lower bound constraint on estimating $\delta$ from the observable is $\sigma^2_{\delta}\geq 1/F^\mathcal{O}_{\delta\delta}$; the equals sign occurs if the error on $\mathcal{O}$ is Gaussian.\footnote{Note that $F^\mathcal{O}_{\delta\delta}$ is unchanged under rescaling of $\mathcal{O}(L)$, i.e. for any arbitrary constant $(A,C)$, $\mathcal{O}(L)$ and $A\mathcal{O}(L)+C$, are equivalent in this context. All the plots of $\mathcal{O}(L)$ shown in the following sections are rescaled arbitrarily for better presentation.}

\subsection{Observing LSSs with an Observable}

To quantify how well an observable measures LSSs, we consider a two-point statistic, the power spectrum of observable map $\hat{\mathcal{O}}$. Since we only consider the power spectrum on large scales, this is equivalent to smoothing fluctuation on the large scale of interest, or the map of $\langle \mathcal{\hat{O}} \rangle$, where $\langle \cdot \rangle$ is the average over many voxels living in the same large-scale $\delta$ value. Since on large scales $\delta \ll 1$, we can linearize $\langle\mathcal{\hat{O}}\rangle$ in $\delta$, and get
\begin{align}
\langle\mathcal{\hat{O}}\rangle(\delta,\mathbf{x}) &=\langle \mathcal{\hat{O}} \rangle (\delta) + \Delta \mathcal{\hat{O}}(\delta,\mathbf{x})\nonumber\\
&=\langle \mathcal{\hat{O}} \rangle (\delta = 0) +\delta\,\partial_\delta \langle \mathcal{\hat{O}} \rangle + \Delta \mathcal{\hat{O}}(\delta,\mathbf{x}).
\end{align}
Here $\mathbf{x}$ refers to the position of the large patch of volume over which the average $\langle \cdot \rangle$ is taken. In this second line, the first term is the fiducial value of $\hat{\mathcal{O}}$, which is a constant across the whole observing volume, so it only contributes to the $k=0$ mode. The second term linearly traces the large-scale overdensity field $\delta$, so it encodes the cosmological clustering information. The last term accounts for the fluctuations due to the Poisson and instrument noise, has no spatial correlation, and thus contributes to the shot noise in the power spectrum. Therefore, the power spectrum consists of the cosmological clustering and shot-noise terms:
\beq
P_{\mathcal{O}}(k) = \left( \partial_\delta \langle \mathcal{\hat{O}} \rangle \right)^2 \, P(k) + P_{\mathcal{O},{\rm shot}},
\eeq
where $P(k)$ is the underlying matter power spectrum and
\beq
P_{\mathcal{O},{\rm shot}} = V_{\rm vox} \, \sigma^2(\mathcal{\hat{O}})
\eeq
is the shot noise, where $\sigma^2(\mathcal{\hat{O}})$ is the variance on $\mathcal{\hat{O}}$ due to the Poisson and instrument noise.  The ratio of the cosmic signal and stochastic noise contributions to the power spectrum can be expressed in terms of the Fisher information $F^{\mathcal{O}}_{\delta \delta}$,
\beq
\frac{\left( \partial_\delta \langle \mathcal{\hat{O}} \rangle \right)^2 \, P(k)}{P_{\mathcal{O},{\rm shot}}}
= \frac{ \left( \partial_\delta \langle \mathcal{\hat{O}} \rangle \right)^2 \, P(k)}{\sigma^2(\hat{\mathcal{O}}) \, V_{\rm vox}} =
\frac{F^{\mathcal{O}}_{\delta \delta}}{V_{\rm vox}} \, P(k).
\eeq
This equation illustrates that it is sufficient to optimize the function $\mathcal{O}(L)$, i.e. to maximize $F^{\mathcal{O}}_{\delta \delta}/V_{\rm vox}$, to minimize the statistical errors in the power spectrum.

In this paper, our goal is to maximize $F^{\mathcal{O}}_{\delta \delta}$ in order to maximize the extracted information from the large-scale density field $\delta$ from a given image (voxel intensity map). This gives the maximum signal-to-noise ratio on the power spectrum of a given image by minimizing the shot noise, and one can use this derived power spectrum to extract the cosmological information. In practice, the optimal observable to constrain $\delta$ might not be the optimal choice for a given specific type of cosmological information. For example, to measure the redshift space distortion, one might prefer an observable that can pick out low-biased tracers to boost the redshift space distortion signals. This practical consideration is beyond the scope of this paper, so we will leave it to the future works.

\subsection{Optimal Observable}

According to \citet{2013MNRAS.434.2961C}, there exists an optimal observable for $\delta$ such that the equality in $F^\mathcal{O}_{\delta\delta}\leq F_{\delta\delta}$ holds;  this observable can extract all the information and give the minimum variance of parameter $\delta$. The optimal observable $\Oopt(L)$ is given by the ``score function'' of parameter $\delta$ evaluated at its fiducial value ($\delta=0$):

\begin{equation}\label{E:def_optobs}
\Oopt(L)=\partial_{\delta} \ln P(L,\delta)|_{\delta=0}.
\end{equation}
This is optimal because its Fisher information is equal to the total Fisher information content per voxel, $F_{\delta\delta}$,
\begin{equation}
F^{\rm opt}_{\delta\delta}=F_{\delta\delta}=\partial_{\delta} \langle \mathcal{\hat{O}}^{\rm opt} \rangle=\langle (\mathcal{\hat{O}}^{\rm opt})^2\rangle
\end{equation}
See Appendix~\ref{A:Fopt_proof} for the proof.

We further define the cumulative optimal Fisher information: 
\begin{equation}\label{E:Fcum}
F_{\delta\delta}^{\rm opt}(L)=\int_{-\infty}^{L'}dL\,P(L')\,(\Oopt)^2(L').
\end{equation}
The limit of $L'\rightarrow\infty$ gives the optimal Fisher information $F^{\rm opt}_{\delta\delta}$. The gradient of $F_{\delta\delta}^{\rm opt}(L)$ is the amount of information gained from each $L$ scale. 

In this work, we are purely concerned with quantifying the (formal) information content. In order to demonstrate the essence of the formalism in the simple and clear context, we will assume some fixed source luminosity function and its response to density field $\delta$, as well as the instrument noise, and quantify the information content under the particular scenario. Therefore, we do not take into account the uncertainties in the modeling of the luminosity function and the relation of the galaxy emission and the underlying density field. 

\section{Toy Model}\label{S:toymodel}

We first start with a toy model to illustrate the concepts introduced above. In this toy model, we assume all the targeted sources have the same luminosity $\ell$, and the luminosity function linearly traces the density field:
\begin{equation}
\Phi(\ell',\delta)=(1+b\,\delta)\,\frac{N}{V_{\rm vox}}\,\delta^D(\ell'-\ell),
\end{equation}
where $\delta^D$ is the Dirac delta function, $N=N(\delta=0)$ is the mean number of sources per voxels, and $b$ is the bias of the source. Here we set $\ell=1$ for convenience.

We further consider a Gaussian noise in the measurement with rms $\sigma_L$, and thus the voxel luminosity pdf reads
\begin{equation}\label{E:toy_PL}
P(L,\delta,\sigma_L)=\sum_{k=0}^{\infty} \frac{e^{-N(\delta)}\,N^k(\delta)}{k!}\,G(L,k\ell,\sigma_L),
\end{equation}
where $N(\delta)=(1+b\,\delta)\,N$ is the expectation value of the number of sources for a voxel residing in density field $\delta$, and 
\begin{equation}
G(x,\bar{x},\sigma)=\frac{1}{\sqrt{2\pi}\,\sigma}e^{-(x-\bar{x})^2/2\sigma^2}
\end{equation}
is the Gaussian function of $x$ with rms $\sigma$ centered at $\bar{x}$.

The GD observable, described by $\OGD(L)$, is a natural choice if $N \ll 1$, so that if a detection is made it is likely coming from a single source, and if $\sigma_L \ll \ell$, so that false detections are unlikely. In this limit, 
the signal is 
\beq
\pa_{\delta} \langle \OGDhat \rangle = b\,N,
\eeq
the (Poisson) variance in $\OGDhat$ reads 
\beq
\sigma^2(\OGDhat) = N,
\eeq
and the Fisher information on the overdensity $\delta$ is
\beq\label{E:poisson_limit}
F^{\rm GD}_{\delta \delta} = b^2N.
\eeq
This is the information on $\delta$ that one obtains from a direct measurement of the number of sources in each voxel, which is only limited by the Poisson noise owing to the finite number of sources, and is thus the maximum attainable information content for a given value of $N$ and $b$. The limit $F=b^2N$ is referred to as the ``Poisson limit'' hereafter. For this reason, below we will compare the ratio, $F/(b^2N)$, of the Fisher information obtained in a given scenario, $F$, to the maximum Fisher information $F = b^2N$.

For IM, the signal is
\beq
\pa_{\delta} \langle \OIMhat \rangle = \pa_{\delta} \langle 
\hat{L}\rangle = \pa_\delta L = b\,N \, \ell,
\eeq
with variance
\beq
\sigma^2(\OIMhat) = \sigma^2(\hat{L}) = \sigma^2_{\rm SN} + \sigma_L^2,
\eeq
where $\sigma^2_{\rm SN} \equiv N \, \ell^2$ is the shot noise due to the finite number of sources contributing to the intensity signal.
This gives
the Fisher information,
\beq
\label{eq:F IM toy}
F^{\rm IM}_{\delta \delta} = \frac{b^2\,N^2 \, \ell^2}{N \, \ell^2 + \sigma_L^2} = b^2N \, \frac{\sigma^2_{\rm SN}}{\sigma^2_{\rm SN} + \sigma^2_L}.
\eeq

In the limit where the noise in the intensity is dominated by the Poisson noise, $\sigma_L \ll \sigma_{\rm SN}$, this gives the optimal result, $F = b^2N$ (Poisson limit). However, in general,
the Fisher information may be suppressed by the instrument noise. If we model variations in voxel volume by changing $N$, Eq.~\ref{eq:F IM toy} shows that the performance of IM as quantified by $F/(b^2N)$ is independent of voxel size as long as either (1) we are in the Poisson-noise-dominated regime $\sigma_L \ll \sigma_{\rm SN}$ or (2) the instrument noise scales with voxel size as $\sigma_L^2 \propto N \propto V_{\rm vox}$. The noise scaling in case (2) is what one would expect if the instrument noise is photon noise dominated.

Below we discuss the optimal observable $\Oopt(L)$, and compare its Fisher information with $\mathcal{O}^{\rm GD}(L)$ and $\mathcal{O}^{\rm IM}(L)$ in three different regimes: $N\ll 1$, $N\sim 1$, and $N\gg 1$.

\subsection{$N\ll 1$}

\begin{figure*}[htbp!]
\begin{center}
\includegraphics[width=\linewidth]{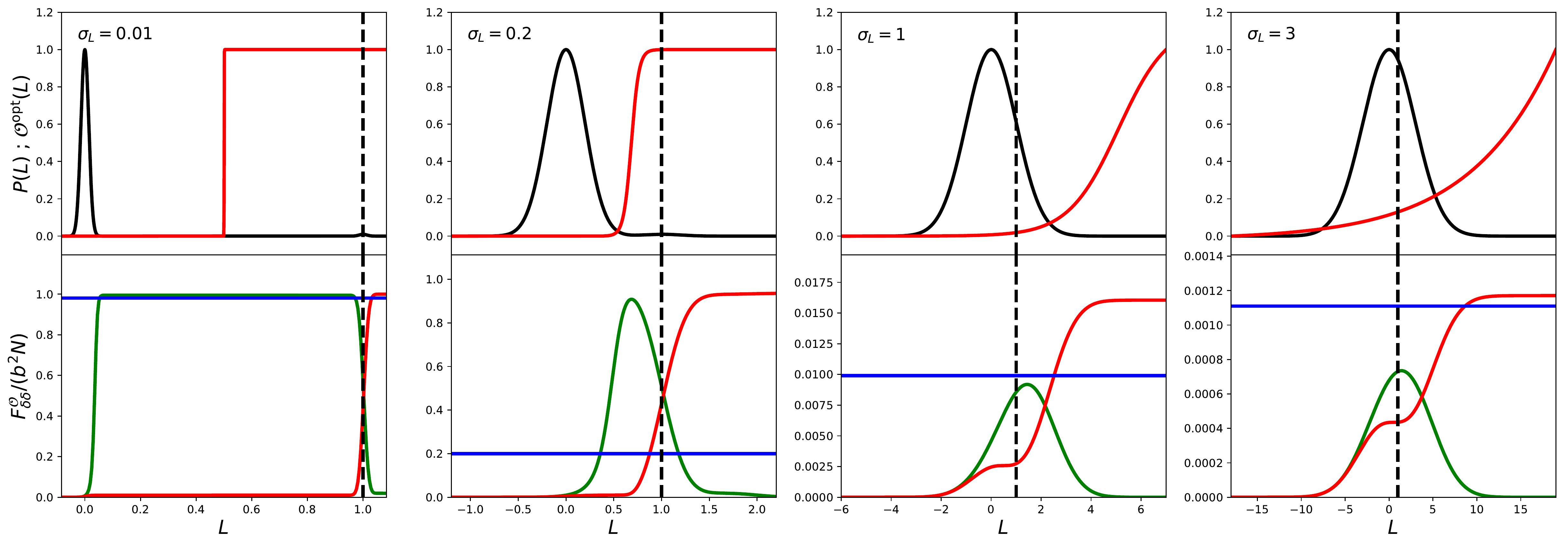}
\caption{\label{F:RdP_Ns}\textbf{Top:}  $P(L)$ (black) and $\Oopt$(L) (red) of the toy model with a single type of source with luminosity $\ell = 1$ and mean number of sources per voxel $N=0.01$, for different Gaussian noise $\sigma_L$. \textbf{Bottom:}  $F^{\mathcal{O}}_{\delta\delta}$ of IM observable (linear function; blue), GD  observables (step function) as a function of step $L$ (green), and the cumulative optimal Fisher information (red). The black dashed lines mark $L=\ell$ for reference.}
\end{center}
\end{figure*}

In the $N \ll 1$ limit, the voxel luminosity probability distribution can be simplified by Taylor-expanding Eq.~\ref{E:toy_PL} and keeping terms only up to first order in $N(\delta)$:
\begin{equation}
P(L,\delta)\simeq (1-N(\delta))\,G(L,0,\sigma_L)+N(\delta)\,G(L,\ell,\sigma_L).
\end{equation}
The optimal observable can then be calculated from Eq.~\ref{E:def_optobs},
\begin{equation}
\Oopt(L)\simeq\frac{b\,N(G(L,\ell,\sigma_L)-G(L,0,\sigma_L))}{(1-N)\,G(L,0,\sigma_L)+N\,G(L,\ell,\sigma_L)}.
\end{equation}

In Fig.~\ref{F:RdP_Ns}, the top panels show $P(L)$ and $\Oopt(L)$, for $N = 0.01$ ($\sigma_{\rm SN}=0.1$) with various instrument noise $\sigma_L$ levels\footnote{The true optimal observable of this case is indeed a stair-like function like the one shown in Fig.~\ref{F:RdP_N1}, rather than a single step we get from approximation with only $k=0,1$ terms. However, this approximation gives almost the same Fisher information as the optimal observable derived from including more $k$ terms. This is due to the fact that the probability of higher $k$ terms is too small to have a significant contribution to Fisher information. Therefore, for the purpose of demonstrating the idea, we ignore the higher-order terms for the optimal observable.}
, and the bottom panels show the Fisher information (See Equation~\ref{E:Fisher}) of the optimal observable (cumulated Fisher information; see Equation~\ref{E:Fcum}), the IM observable, and the GD observable for a range of threshold $L_{\rm th}$.

Considering first the low-noise regime, $\sigma_L \ll \ell$ (left panels), we find as expected that thresholded GD is optimal. This is clearly seen from the fact that the optimal observable $\Oopt(L)$ (red curve) is close to a step function. In addition, the Fisher information of $\OGD(L)$ as a function of $L_{\rm th}$ attains approximately the same total information as the optimal observable, for a wide range of values of $L_{\rm th}$. Any threshold from a few times $\sigma_L$ to $\ell$ minus a few times $\sigma_L$ perfectly ``counts'' sources. As a result, the information content is optimal, in the sense that $F/(b^2N) = 1$. 

In the very low noise regime, $\sigma_L \ll \sigma_{\rm SN}$ (where $\sigma_{\rm SN}$ is the Poisson noise in luminosity $L$), IM is {\it also} optimal, as can be seen by the horizontal blue line in the bottom panel. This is because in the $N\ll 1$ and low-noise ($\sigma_L^2 \ll N\ell^2$) limit, most voxels have either $L\approx 0$ or $L\approx \ell$, as shown by the $P(L)$ function, and thus the information content must be concentrated at these two $L$ scales as well. As long as an observable is able to discriminate these two classes of voxels, i.e. having distinct values at $L=0$ and $L=\ell$, it is able to capture the signals (quantified by $\partial_\delta \langle \mathcal{\hat{O}}\rangle$) in the map, regardless of the $\mathcal{O}$(L) function values at other $L$ values, as almost no voxel falls in this regime. However, in the intermediate regime ($\sigma_L=0.2$ case), $\sigma_{\rm SN} < \sigma_L \ll \ell$, IM suffers from instrument noise suppression (see Equation~\ref{eq:F IM toy}), while source detection is still optimal.

Moving on from the low-noise regime toward cases where $\sigma_L \ll \ell$ no longer holds ($\sigma_L=1,3$), the Gaussian noise profiles of the $P(L)$ function centered at 0 and $\ell$ start to overlap, so a GD threshold function is no longer optimal, as it cannot effectively count the sources. Indeed, the optimal observable $\Oopt(L)$ is now a more gradually increasing function of $L$. As for the Fisher information, we can see from Fig.~\ref{F:RdP_Ns} that even for the optimal choice of $L_{\rm th}$, the information contained in the GD observable is lower than the information in the optimal observable. At the same time, the IM information content becomes larger relative to the optimal information content. In the largest noise regime ($\sigma_L=3$), IM is very close to optimal.

We note, however, that as the noise increases, the {\it absolute} information content strongly decreases, i.e., $F/(b^2N) \ll 1$. This is of course to be expected: instrument noise makes it difficult to measure cosmological signals.

\subsection{$N\sim 1$}

\begin{figure*}[htbp!]
\begin{center}
\includegraphics[width=\linewidth]{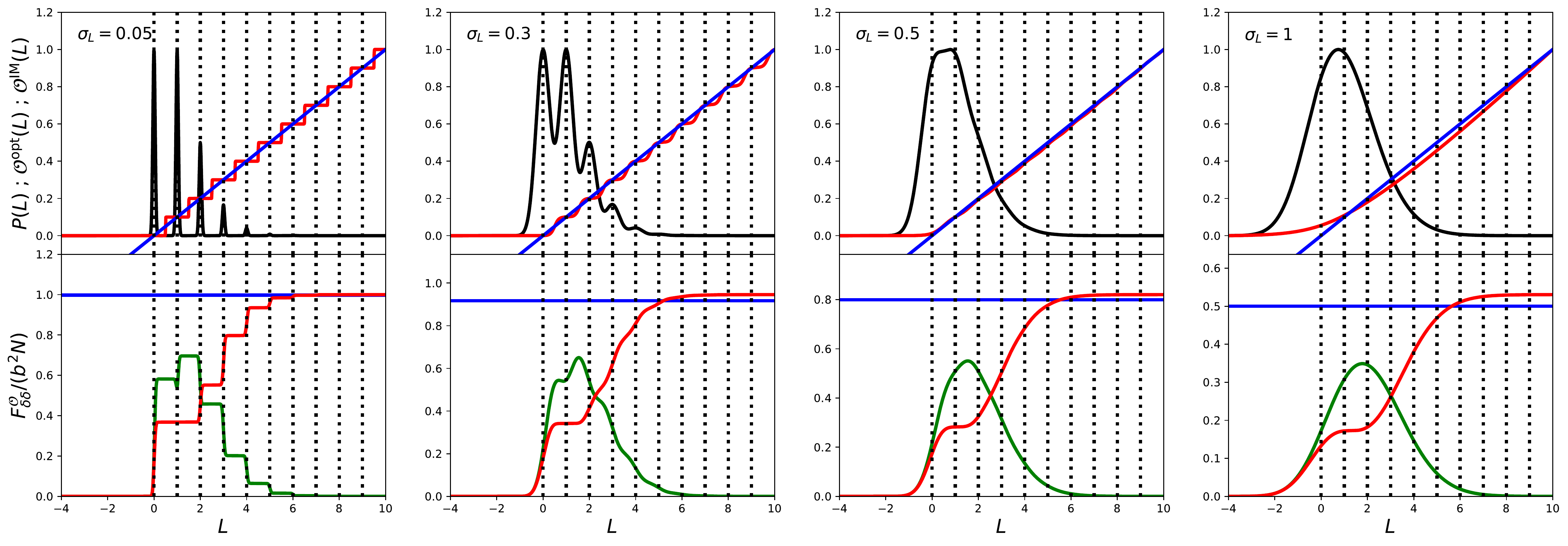}
\caption{\label{F:RdP_N1}\textbf{Top:} $P(L)$ (black) and $\Oopt$(L) (red) of toy model with a single type of source with luminosity $\ell = 1$ and mean number of sources per voxel $N=1$, for different Gaussian noise $\sigma_L$. \textbf{Bottom:}  $F^{\mathcal{O}}_{\delta\delta}$ of IM observable (linear function; blue), GD  observables (step function) as a function of threshold $L$ (green), and the cumulative optimal Fisher information (red). The black dotted lines mark the integer of $\ell$, the possible intrinsic voxel luminosity.}
\end{center}
\end{figure*}

Next, we consider the $N\sim 1$ regime. In this scenario, the $k\geq 2$ terms in Eq.~\ref{E:toy_PL} must be taken into account. We take $N=1$ in this example and consider different $\sigma_L$ values as before. The results are shown in Fig.~\ref{F:RdP_N1}. The $P(L)$ function is the linear combination of the Gaussian profile with variance $\sigma_L^2$ centered at $L=0,\,\ell,\,2\ell,...$, with their amplitude following a Poisson distribution. We can see that the optimal observable is a stair-like function, which gradually smoothed out with increasing noise. 

The linear observable is better than the step function in all cases in terms of their Fisher information. The reason is the same as in the $N\ll 1$ situation: in the low-noise regime, where most voxel luminosity $L$ has values around $L=0,\,\ell,\,2\ell,...$, the only observable value that matters is where $L$ is near these values. The linear observable gives exactly the same value at these points as the optimal one. On the other hand, the step function is not a good observable in this case. The step function gives the same weights for all the voxels above the step, so it ignores the fact that higher-luminosity voxels likely have more sources and are more likely to reside in high-$\delta$ regions. Note that this is not an issue for the $N\ll1$ case, as there are very few voxels containing multiple sources; the total information content in these voxels is also negligible. Whereas here we have $N\sim 1$, the multiple-source voxels contribute to a significant portion of the total information content, and a proper weighting for them in the observable is essential for capturing the information from the map.

In the high instrument noise regime, the linear observable is also superior to the step function, which follows the same argument as in the $N\ll 1$ case.

\subsection{$N\gg 1$}\label{S:toyNl}

\begin{figure*}[htbp!]
\begin{center}
\includegraphics[width=\linewidth]{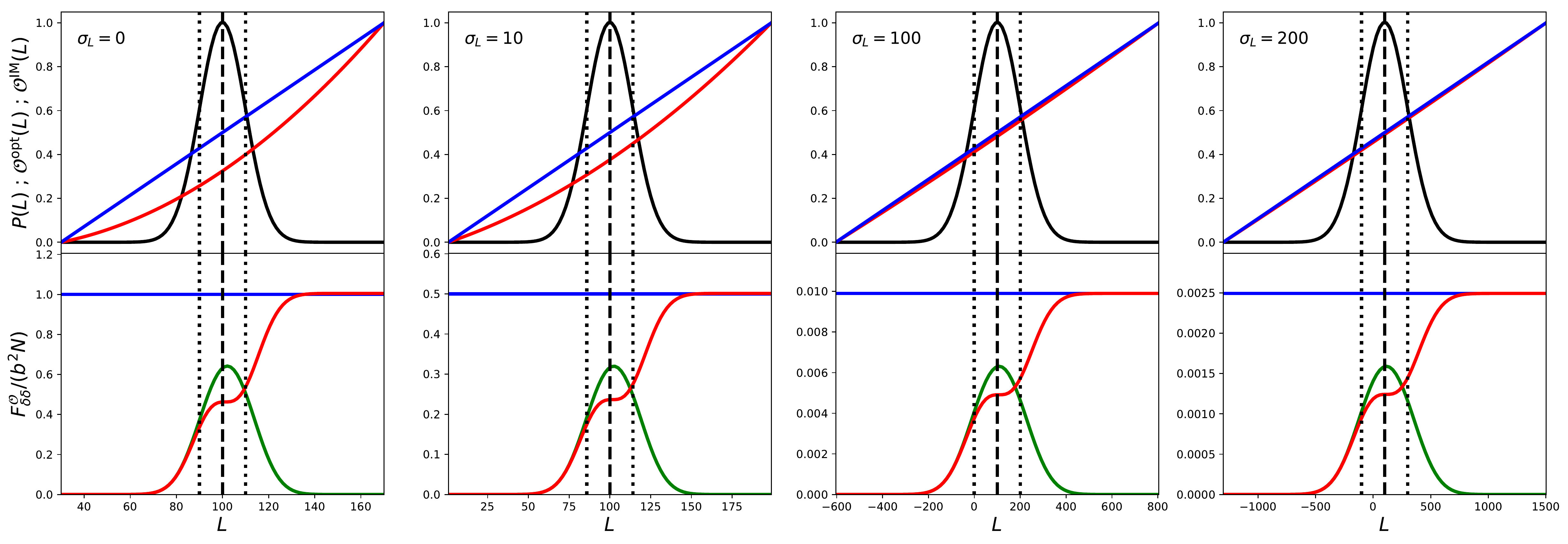}
\caption{\label{F:RdP_Nl}\textbf{Top:} $P(L)$ (black) and $\Oopt$(L) (red) of the toy model with a single type of source with luminosity $\ell = 1$ and mean number of sources per voxel $N=100$, for different Gaussian noise $\sigma_L$. \textbf{Bottom:}  $F^{\mathcal{O}}_{\delta\delta}$ of IM observable (linear function; blue), GD  observables (step function) as a function of step $L$ (green), and the cumulative optimal Fisher information (red). The black dashed and dotted lines mark the mean and $\pm\bar{\sigma}$ of the $P(L)$ profile, i.e. $N\ell$ and $N\ell\pm \bar{\sigma}$, respectively.}
\end{center}
\end{figure*}

In the $N\gg 1$ limit, the Poisson function converges to a Gaussian,
\begin{equation}
\frac{e^{-N(\delta)}\,N^k(\delta)}{k!}\simeq \frac{1}{\sqrt{2\pi\,N(\delta)}}\,e^{-\frac{(N(\delta)-k)^2}{2\, tN(\delta)}}
\end{equation}
and the summation over $k$ in the $P(D)$ formalism can be approximated by an integral, so Eq.~\ref{E:toy_PL} becomes the convolution of two Gaussian functions, which gives another Gaussian,
\begin{equation}
\begin{split}
&P(L,\delta)=\sum_{k=0}^{\infty}\, \frac{e^{-N(\delta)}\,N^k(\delta)}{k!}\,G(L,k\ell,\sigma_L)\\
&\simeq \int_{0}^{\infty}dk \left [ \frac{1}{\sqrt{2\pi\,N(\delta)}}e^{-\frac{(k-N(\delta))^2}{2\,N(\delta)}} \right ]\left [ \frac{1}{\sqrt{2\pi}\,\sigma_L}e^{-\frac{(L-k\ell)^2}{2\,\sigma_L^2}} \right ] \\
&=\frac{1}{\sqrt{2\pi}\,\bar{\sigma}}e^{-\frac{L'^2}{2\,\bar{\sigma}^2}},
\end{split}
\end{equation}
where $L'(\delta)\equiv L-N(\delta)\,\ell$, and $\bar{\sigma}^2(\delta)\equiv\sigma_L^2+N(\delta)\,\ell^2$. Note that $\bar{\sigma}^2$ is the total variance from both instrument noise and Poisson noise. In the absence of instrument noise, we still have a nonzero voxel pdf P(L) owing to the Poisson variance of the sources themselves. We then derive the optimal observable from Eq.~\ref{E:def_optobs}, with some rescaling to get rid of all irrelevant constants,\footnote{$L'\equiv L'(\delta =0) = L-N\,\ell$; $\bar{\sigma}^2 \equiv \bar{\sigma}^2(\delta=0)=\sigma_L^2+N\,\ell^2$}
\begin{equation}\label{E:RdP_Nl_oopt}
\Oopt(L)=L'+\frac{\ell}{2\,\bar{\sigma}^2}\,L'^2.
\end{equation}
Hence, the optimal observable is a linear combination of a linear and a quadratic term, and the contribution from the latter gets smaller as the noise increases.

The top row of Fig.~\ref{F:RdP_Nl} shows the $P(L)$ and $\Oopt(L)$ for different $\sigma_L$ levels, while fixing $N=100$. We can see that as $\sigma_L$ increases, the $P(L)$ profile is broadened, and $\Oopt(L)$ becomes closer to the linear function. The bottom row shows the Fisher information for the different observables. In all cases the step function is not the preferable observable. The linear function performs as well as the optimal observable, even in the $\sigma_L=0$ limit, where the optimal observable deviates from the linear function significantly. This is because the quadratic term in the optimal observable has negligible contribution to the optimal Fisher information (see Appendix.~\ref{A:lin_quad} for explanation).

\subsection{Toy Model Summary}

In conclusion, for our toy model with a luminosity function describing sources with a single luminosity $\ell$, we find the following limiting behaviors:
\begin{itemize}
\item For a low number of sources per voxel, $N \ll 1$, {\it and} low noise compared to the source luminosity, $\sigma_L \ll \ell$, it is optimal to detect individual sources by applying the threshold observable $\mathcal{O}^{\rm GD}(L)$. In this scenario, the voxels below the detection threshold contain only noise and make up the majority of voxels. The GD observable assigns them zero weight, and therefore they do not contribute to the noise in the map. On the other hand, voxels with luminosity above the threshold all contain a (single) source (as the probability of a noise fluctuation exceeding the threshold is infinitesimally small in the limit $\sigma_L \ll \ell$). This leads to a measurement of the source number density only limited by the shot noise owing to the finite number of sources $N$.
\item In the same low-$N$ but high-noise regime where $\sigma_L > \ell$, the signal from sources cannot be unambiguously distinguished from noise fluctuations, so that the GD approach is suboptimal and instead the IM observable is close to optimal. The measurement is limited by instrument noise (as opposed to by shot noise owing to the finite number of sources), so that our ability to constrain $\delta$ (as quantified by the Fisher information) is unsurprisingly much weaker than the one in the $\sigma_L \ll \ell$ regime.
\item In the opposite regime of a large number of sources per voxel, $N \gg 1$, we find that IM is (nearly) optimal independently of the instrument noise.
\end{itemize}

The above results are intuitive and serve as useful benchmarks to refer to in the following sections. Intermediate cases can be understood as interpolations between the above limiting scenarios.

\section{Schechter Luminosity Function Model}\label{S:SchLF} 
For a more realistic description, we consider taht the galaxy populations follow a Schechter luminosity functional form: $\Phi(\ell)=\phi_\ast\, (\ell/\ell_\ast)^\alpha \,e^{-\ell/\ell_\ast}$\citep{1976ApJ...203..297S}\footnote{To simplify the notations, $\Phi(\ell)$ refers to $\Phi(\ell,\delta=0)$, the average luminosity function across the universe.}. To simplify the notation, below all the $\ell$ represent $\ell/\ell_\ast$; in other words, we use $\ell_\ast$ as the unit for luminosity. This can be easily scaled to any desired unit in real experiments. 

One requirement for applying the $P(D)$ formalism is to have a finite ${N}$, the mean number of sources per voxel. To ensure that the integration in Eq.~\ref{E:Ndef} converges, we use a modified Schechter function introduced by \citet{2017MNRAS.467.2996B}
\begin{equation}\label{E:lfunc}
\Phi(\ell)=\phi_\ast\, \ell^\alpha\, e^{-\ell}\,e^{-\ell_{\rm min}/\ell}.
\end{equation}

We assume that the luminosity function linearly traces the density field, 
\begin{equation}
\label{eq:deltag effect}
\Phi(\ell,\delta)=(1+b\,\delta)\,\Phi(\ell).
\end{equation}
The optimal observable, $P(L)$, and $F^{\mathcal{O}}_{\delta\delta}$ can be derived from equations in Sec.~\ref{S:formalism}. Note that Eq.~\ref{eq:deltag effect} assumes a luminosity-independent clustering bias. In a more realistic description, we would describe the response to the underlying {\it matter} overdensity $\delta$ in terms of a luminosity-dependent bias $b(\ell)$. This is a straightforward modification to our formalism, but for simplicity we will not pursue it here.

Applying the low-$\ell$ suppression for $\ell\lesssim\ell_{\rm min}$ has a physical motivation: galaxies cannot be infinitely faint. The value of $\ell_{\rm min}$ is not easily constrained observationally; however, it is not an issue for our calculation. In Appendix~\ref{A:Lmin}, we show that the choice of $\ell_{\rm min}$ does not affect our results as long as $\ell_{\rm min}$ is much smaller than $\sigma_L$, the instrumental noise in the observation. In this work, we adopt the fiducial $\ell_{\rm min}=10^{-3}$.

The faint-end slope $\alpha$ usually has the value $-2<\alpha<-1$ from observations. We take $\alpha=-1.5$ as our fiducial value in this work, and we discuss the effects of choosing different $\alpha$ values in Appendix~\ref{A:alpha}.

\subsection{Quantifying the Confusion}

\begin{figure}[htbp!]
\begin{center}
\includegraphics[width=\linewidth]{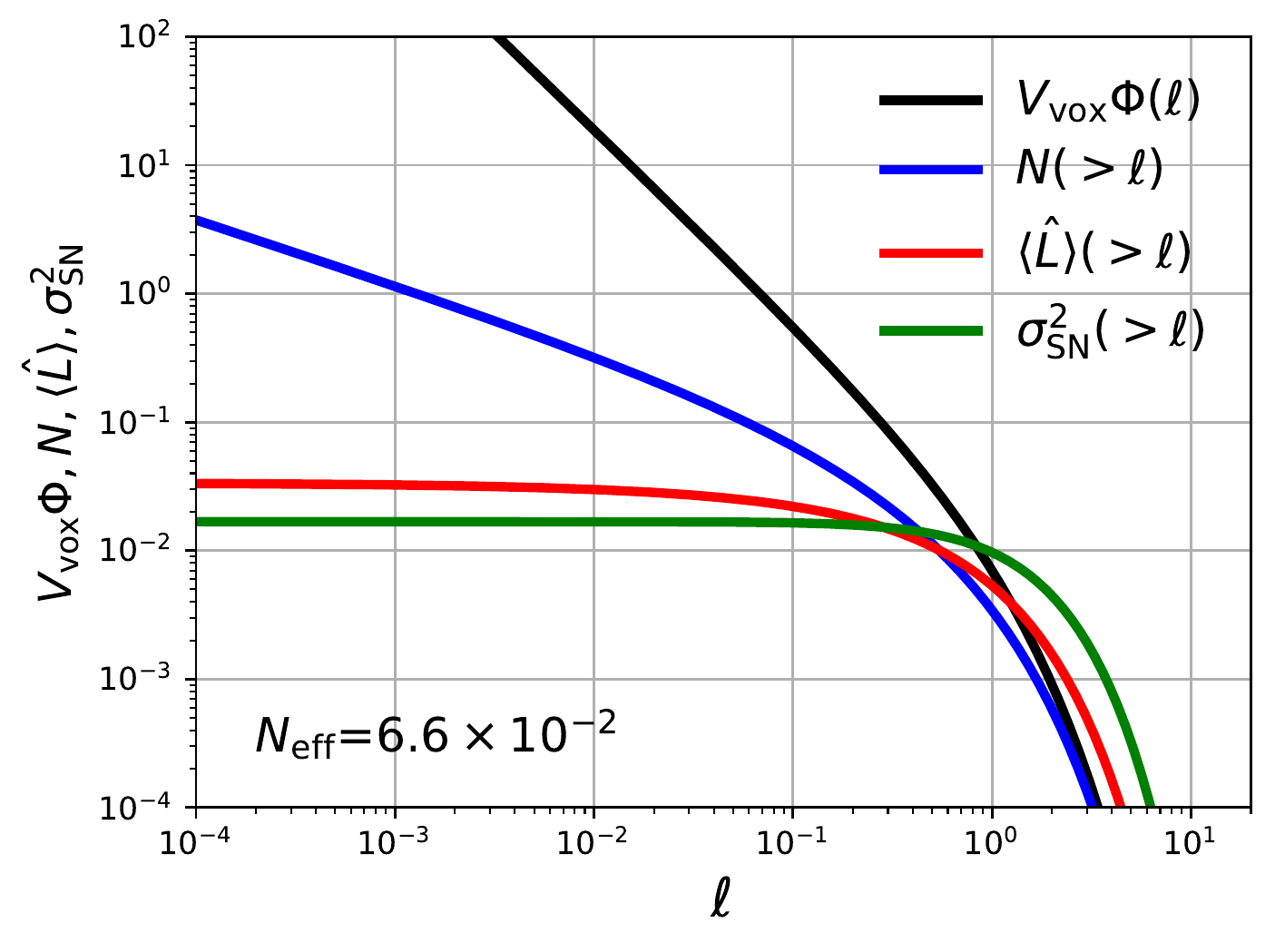}
\caption{\label{F:Nint} Normal Schechter luminosity function (without $\ell_{\rm min}$ cutoff) using fiducial $\alpha=-1.5$ (black), and its cumulative $N$ (blue), $\langle\hat{L}\rangle$ (red), and $\sigma_{\rm SN}^2$ (green).}
\end{center}
\end{figure}

 Fig.~\ref{F:Nint} shows the normal Schechter function (without $\ell_{\rm min}$ cutoff) with fiducial $\alpha$. We also plot the first three moments of the Schechter function that give the quantity of particular interest: 
 \begin{align}
 N&=V_{\rm vox}\int d\ell\,\Phi(\ell) \\
 \langle\hat{L}\rangle&=V_{\rm vox}\int d\ell\,\Phi(\ell)\,\ell\\
 \sigma_{\rm SN}^2&=V_{\rm vox}\int d\ell\,\Phi(\ell)\,\ell^2
 \end{align}

As shown in the plot, the total number of sources $N$ diverges as we take $\ell_{\rm min}$ to zero, corresponding to an infinite number of (mostly faint) sources per voxel in the absence of a cutoff. As a result, the value of $N$ in the modified Schechter function depends on the choice of $\ell_{\rm min}$, while for $\langle\hat{L}\rangle$ and $\sigma^2_{\rm SN}$, the integration is converged at the faint end, so its value is not susceptible to the artificial $\ell_{\rm min}$ cutoff (these convergence properties are true for all $-2<\alpha<-1$).
 
For the above reasons, $N$ is not a well-defined quantity in the Schechter function case and is ill-suited to quantify the level of confusion as used in the toy model. We therefore introduce an effective number of sources per voxel, $N_{\rm eff}$, defined with the cutoff-independent quantities $\langle\hat{L}\rangle$ and $\sigma_{\rm SN}^2$.

\subsubsection{$N_{\rm eff}$}

The IM signal in the Schechter model is given by
\beq
\pa_\delta \langle \OIMhat \rangle = \pa_\delta \langle \hat{L} \rangle = b\,V_{\rm vox} \, \int d\ell \, \Phi(\ell) \, \ell,
\eeq
with variance
\beq
\sigma^2(\OIMhat) = \sigma^2(\hat{L}) = \sigma^2_{\rm SN} + \sigma^2_L.
\eeq
The Fisher information is therefore
\beq
\label{eq:FIM schechter}
F^{\rm IM}_{\delta \delta} = \frac{b^2\left( V_{\rm vox} \, \int d\ell \, \Phi(\ell) \, \ell \right)^2}{V_{\rm vox} \, \int d\ell \, \Phi(\ell) \, \ell^2 + \sigma_L^2}=\frac{b^2\langle\hat{L}\rangle^2}{\sigma_{\rm SN}^2+\sigma_L^2}.
\eeq

We now define the effective number of sources per voxel as the IM Fisher information in the
Poisson-limited case, $\sigma_L \ll \sigma_{\rm SN}$ ,
\beq
N_{\rm eff} \equiv \frac{\left( V_{\rm vox} \, \int d\ell \, \Phi(\ell) \, \ell \right)^2}{V_{\rm vox} \, \int d\ell \, \Phi(\ell) \, \ell^2}=\frac{\langle\hat{L}\rangle^2}{\sigma_{\rm SN}^2}.
\eeq
This can be interpreted as the reciprocal of the effective shot noise in the IM regime, which is an analogy to the $1/N$ shot noise in GD. 

The total Fisher information from IM (Eq.~(\ref{eq:FIM schechter})) can be rewritten as
\beq
\label{E:F_IM_Sch}
F^{\rm IM}_{\delta \delta} = b^2N_{\rm eff} \, \frac{\sigma^2_{\rm SN}}{\sigma^2_{\rm SN} + \sigma^2_L}.
\eeq
The effective number of sources per voxel thus tells us how well the IM observable can possibly perform given a source population, while the performance is weakened when $\sigma_L \gtrsim \sigma_{\rm SN}$. As is the case for the toy model, the IM performance is independent of $V_{\rm vox}$ if the instrument noise scales like $\sigma_L^2 \propto V_{\rm vox}$ or if the instrument noise is negligible, $\sigma_L \ll \sigma_{\rm SN}$.

\subsubsection{$L_{\rm SN}$}

\begin{figure}[htbp!]
\begin{center}
\includegraphics[width=\linewidth]{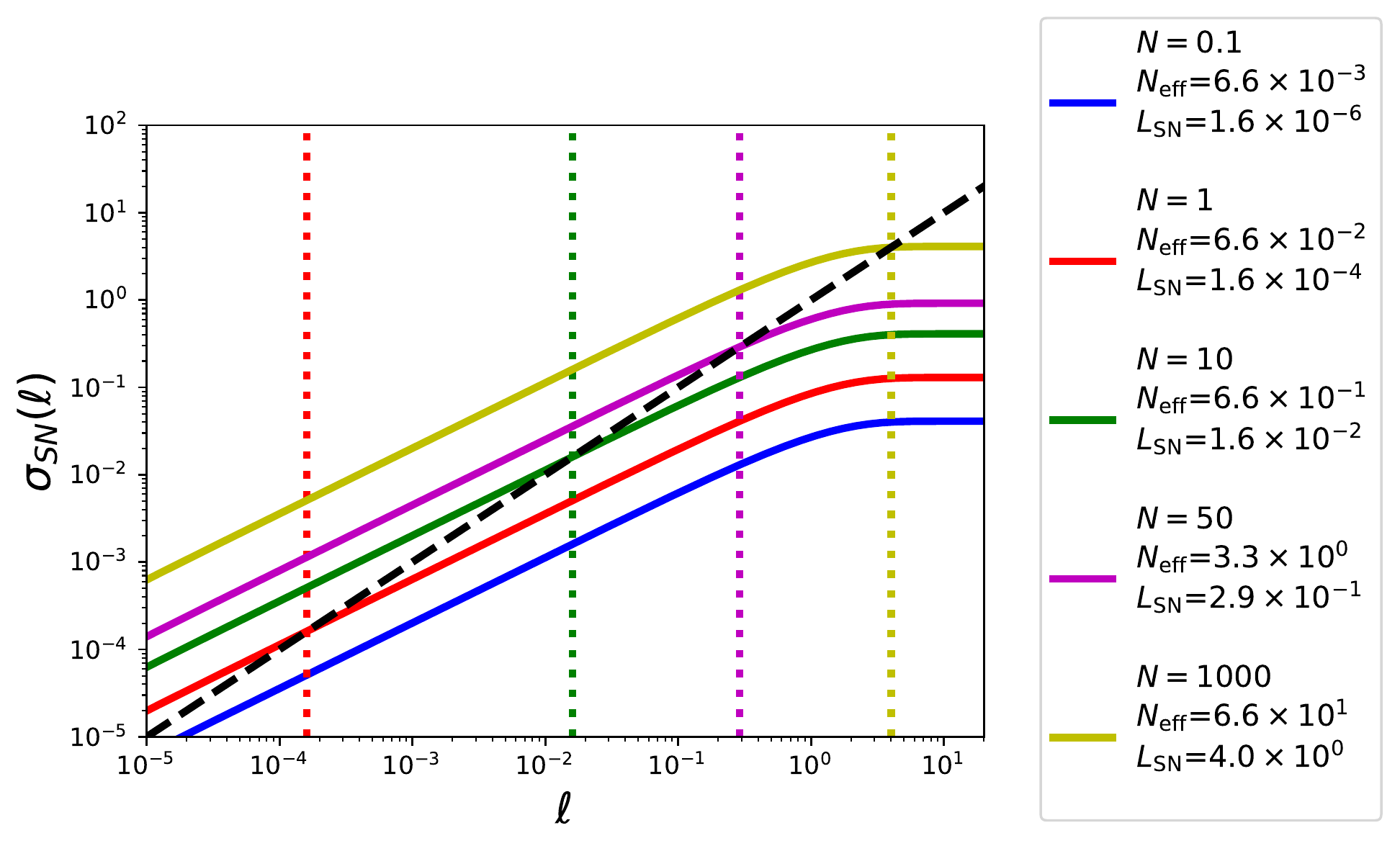}
\caption{\label{F:sigsn} $\sigma_{\rm SN}(\ell)$ with different source densities (solid lines). The black dashed line is $\sigma_{\rm SN}=\ell$, and its intersection with $\sigma_{\rm SN}(\ell)$ is $L_{\rm SN}$.}
\end{center}
\end{figure}

Aside from $N_{\rm eff}$, we further introduce the luminosity scale where the voxels are highly susceptible to shot noise, $L_{\rm SN}$, to be another quantity related to confusion.

We first define the cumulative intensity shot noise,
\begin{equation}\label{E:sig_SN}
\sigma_{\rm SN}^2(\ell) \equiv V_{\rm vox} \, \int_0^{\ell}d\ell' \,  \Phi(\ell') \, \ell'^2.
\end{equation}
This includes the shot-noise variance from all the sources fainter than $\ell$. A useful quantity is then the ``crossover luminosity,'' $L_{\rm SN}$, where the intensity shot noise equals the source luminosity,
$\sigma_{\rm SN}(L_{\rm SN}) = L_{\rm SN}$.

When $\ell< L_{\rm SN}$, $\sigma_{\rm SN}(\ell)> \ell$, which means that the confusion noise from the fainter source is comparable to $\ell$; when $\ell> L_{\rm SN}$, $\sigma_{\rm SN}(\ell)< \ell$, which means that the confusion noise from faint sources becomes negligible.  Fig.~\ref{F:sigsn} shows the $\sigma_{\rm SN}(\ell)$ with four different source densities and their $L_{\rm SN}$ marked by the dotted vertical lines. 

\subsubsection{Relation between $N_{\rm eff}$ and $L_{\rm SN}$}\label{S:NeffLsn}
The modified Schechter luminosity function we adopted in this work is composed of a power law with slope $\alpha$ and exponential cutoffs at both low- and high-$\ell$ ends, which guarantee convergence of integration for all moments. Of particular interest are the first three moments that give $N$ (zeroth), $\langle\hat{L}\rangle$ (first), $\sigma_{\rm SN}^2$ (second) respectively. 

If the luminosity function is only a power law (i.e. $\Phi\propto\ell^{\alpha}$) with $-2<\alpha<-1$ , the zeroth moment converges at the high-$\ell$ end and diverges at the low$-\ell$ end, while the convergence of higher moments is reversed. Applying the exponential cutoff suppresses contribution from scales beyond the cutoff scale, and thus the integration is dominated by the sources with luminosity around the cutoff. Therefore,
\begin{align}
N&=V_{\rm vox}\int\Phi(\ell)\,d\ell\sim V_{\rm vox}\,\Phi(\ell_{\rm min})\,\ell_{\rm min}\\
\langle \hat{L} \rangle&=V_{\rm vox}\int\Phi(\ell)\,\ell\, d\ell\sim V_{\rm vox}\,\Phi(\ell_\ast)\,\ell_\ast^2\\
\sigma_{\rm SN}^2&=V_{\rm vox}\int\Phi(\ell)\,\ell^2\,d\ell\sim V_{\rm vox}\,\Phi(\ell_\ast)\,\ell_\ast^3.
\end{align}

Note that the quantity $\ell\,\Phi(\ell)$ is the count per log $\ell$, so the above approximations imply that $N$ is dominated by sources with luminosity around $\ell_{\rm min}$, whereas $\langle \hat{L} \rangle$ and $\sigma_{\rm SN}^2$ are dominated by $\ell \sim\ell_*$ sources.

From these relations we can also derive
\begin{equation}
N_{\rm eff}=\frac{\langle\hat{L}\rangle^2}{\sigma_{\rm SN}^2} \sim V_{\rm vox}\,\Phi(\ell_\ast)\,\ell_\ast, 
\end{equation}
so $N_{\rm eff}$ is approximately the number of sources per log$(\ell)$ at $\ell_\ast$.

Based on the above, we can roughly infer the relation between $L_{\rm SN}$ and $N_{\rm eff}$. Since 
\begin{equation}
\sigma^2_{\rm SN}(L_{\rm SN})\equiv L_{\rm SN}^2 \sim V_{\rm vox}\,\Phi(L_{\rm SN})\,L_{\rm SN}^3,
\end{equation}
if $L_{\rm SN}<\ell_\ast$, we get 
\begin{equation}
V_{\rm vox}\,\Phi(L_{\rm SN})\,L_{\rm SN}\sim 1 >  V_{\rm vox}\,\Phi(\ell_*)\,\ell_\ast=N_{\rm eff}.
\end{equation}
On the contrary, if $L_{\rm SN}>\ell_\ast$, then
\begin{equation}
V_{\rm vox}\,\Phi(L_{\rm SN})\,L_{\rm SN}\sim 1 <  V_{\rm vox}\,\Phi(\ell_*)\,\ell_\ast=N_{\rm eff}.
\end{equation}

\vskip 7pt

Hence, we conclude that 
\begin{align}
&L_{\rm SN}<\ell_\ast \,\Leftrightarrow \, N_{\rm eff}<1\nonumber\\
&L_{\rm SN}>\ell_\ast \,\Leftrightarrow \, N_{\rm eff}>1
\end{align}

\vskip 7pt

\begin{figure}[htbp!]
\begin{center}
\includegraphics[width=\linewidth]{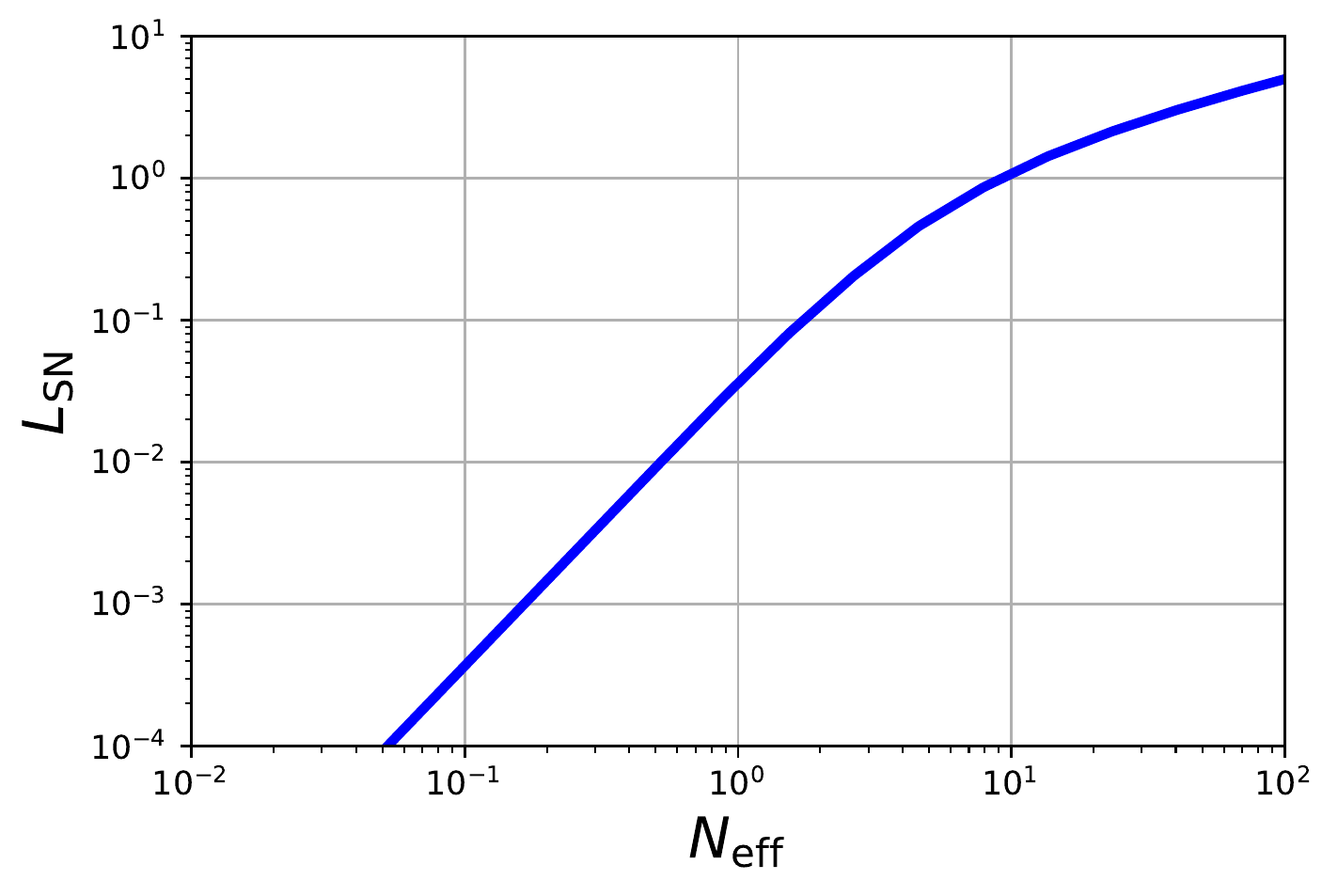}
\caption{\label{F:LsnNeff} $L_{\rm SN}-N_{\rm eff}$ relation with fiducial Schechter function faint-end slope $\alpha=-1.5$. Note that the actual scales where $N_{\rm eff}=1$ and $L_{\rm SN}=\ell_*(=1)$ happen are off by around an order of magnitude.}
\end{center}
\end{figure}

The argument above is only an order-of-magnitude estimation. The $L_{\rm SN}-N_{\rm eff}$ relation with our fiducial Schechter parameters is shown in Fig.~\ref{F:LsnNeff}. The actual scales where $N_{\rm eff}=1$ and $L_{\rm SN}=\ell_*(=1)$ happen are off by around an order of magnitude. Later we will focus on the limiting scenarios where $L_{\rm SN}\ll \ell_*$ and $L_{\rm SN}\gg \ell_*$ respectively. In the situation where $L_{\rm SN}\sim \ell_*$ within roughly an order of magnitude, one should keep in mind the caveat that the cases of interest might be closer to either of the limiting regimes, or some intermediate situation, so the arguments for the limiting cases cannot be applied naively.

\subsection{Noiseless Scenario}
We first consider an idealized scenario without instrument noise $\sigma_L$. This example will allow us to derive some useful insights before we move on to the more realistic scenario including instrument noise $\sigma_L$.

The major difference between the toy model and the Schechter function case is that in the toy model with zero instrumental noise, even in the highly confused scenario ($N\gg 1$), the Fisher information of the optimal observable (and of $\OIM(L)$) still reaches the Poisson limit, since we can unambiguously count the number of sources for any given voxel luminosity $L$ in the toy model. In the Schechter function case, on the other hand, we are not able to distinguish the exact composition of sources in the voxels, and thus the information content will be suppressed by the confusion.

\begin{figure}[htbp!]
\begin{center}
\includegraphics[width=\linewidth]{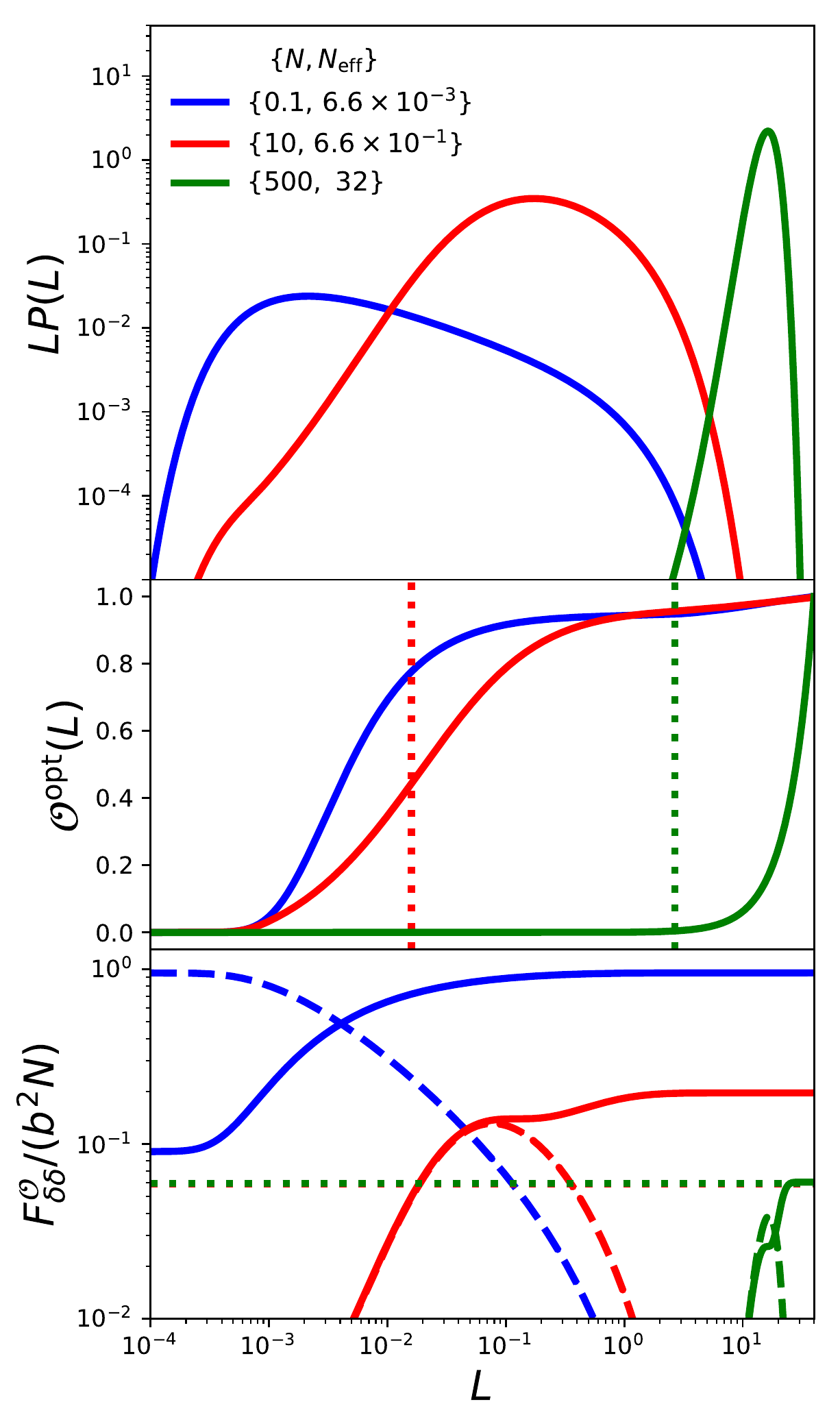}
\caption{\label{F:A_noiseless}Fiducial Schechter function without instrument noise. \textbf{Top:} $LP(L)$ with different $N$ levels. Note that the area underneath the $LP(L)$ curve gives the probability per log $L$. \textbf{Middle:} The optimal observables for each case. The dotted lines mark the $L_{\rm SN}$ (in $N=0.1$ case, $L_{\rm SN}\ll 10^{-4}$, so the blue dot line is outside the x-axis range). \textbf{Bottom:} $F^{\mathcal{O}}_{\delta\delta}$ of IM observable (dotted; note that the three dotted lines overlap), GD  observable as a function of step $L$ (dashed), and the cumulative optimal Fisher information (solid). }
\end{center}
\end{figure}

Fig.~\ref{F:A_noiseless} shows the $P(L)$, $\Oopt(L)$, and the Fisher information relative to the total information from directly counting sources, $F/(b^2N)$, for three different $N$ levels. Below we describe the important observations from these results. 

\begin{itemize}
\item 
The probability distribution of the total voxel luminosity, $P(L)$, shifts to higher $L$ as $N$ increases. 

\item 
The optimal observable has a smoothed step-function-like shape. The transition $L$ scale is around $L_{\rm SN}$, except for the $N=0.1$ case, where $L_{\rm SN}\ll \ell_{\rm min}$, and the transition is strongly affected by the cutoff $\ell_{\rm min}$. The interpretation is as follows: when $L\lesssim L_{\rm SN}$, $\sigma_{\rm SN}\gtrsim L$ (and the effective number of sources below $L$ is not small), and thus the possibility that a given $L$ voxel is composed of multiple faint sources is non-negligible. In this regime, the optimal observable prefers giving brighter voxels more weight since they are more likely to hold more sources, and this explains the rising part of the $\mathcal{O}^{\rm opt}$ function. On the bright end, where $L>L_{\rm SN}$, most of the voxels with these $L$ values are dominated by the single $\ell\sim L$ source, and thus this is in the GD regime, and the optimal observable is a uniform weighting. 

\item 
The $N=0.1$ case reaches the Poisson limit. This is because a threshold $L_{\rm th}$ below $\ell_{\rm min}$ has the property that whenever a voxel luminosity exceeds $L_{\rm th}$, that voxel is likely to contain only a single source. Thus, (only) this scenario allows us to directly count galaxies and thus to optimally trace the overdensity $\delta$. For larger $N$, only sources with $\ell > L_{\rm SN} > \ell_{\rm min}$ can be ``counted.''

\item
In the $N=0.1$ case, the step function with threshold $L_{\rm th} < \ell_{\rm min}$ is approximately optimal as discussed above.

\item
In the two larger-$N$ scenarios, the confusion has a significant impact on fainter voxels ($L\lesssim L_{\rm SN}$) that degrades the information content, and thus the optimal Fisher information is less than the Poisson limit.

\item 
In the two larger-$N$ scenarios, the optimal Fisher information is built up at two stages that correspond to the IM part at $L\lesssim L_{\rm SN}$, where the observable is weighted by luminosity, and the GD part at $L\gtrsim L_{\rm SN}$, where the bright sources can be counted individually. 

\item
In the absence of instrument noise, $F^{\rm IM}_{\delta\delta}/(b^2N)$ is independent of $N$ (and thus the voxel size). This can be understood in the following way: the IM observable measures a luminosity-weighted ``count'' of the number of sources. Because of the properties of the Schechter function discussed in Sec.\ref{S:NeffLsn}, this weighted count is dominated by sources with luminosity near $\ell_*$ ($=1$), and the information content is given by $N_{\rm eff} \ll N$. See also Appendix.~\ref{A:noiseless_lin} for further discussion of this point. 
\end{itemize}

In summary, when $N$ is not small, confusion, in combination with  a range of source luminosities, implies that we cannot reach the Poisson limit even without instrument noise. The IM observable never reaches the Poisson limit, regardless of $N$, while GD reaches $F/(b^2N) = 1$ only if $N \ll 1$. 

\subsection{General Case with Instrumental Noise $\sigma_L$}\label{S:Sch_noise}

In reality, the instrumental noise $\sigma_L$ has to be taken into account. Just as $L_{\rm SN}$ sets the approximate luminosity where a source rises above the confusion noise due to fainter objects, $\sigma_L$ determines the luminosity where objects rise above the instrument noise. Another characteristic scale is the $\ell_\ast$ of the Schechter function, which is set to unity in this paper as we scale luminosities in units of $\ell_\ast$. The shape of the optimal observable and the Fisher information are determined by the relative value of these three luminosity scales $\{L_{\rm SN}, \sigma_L, \ell_*\}$. In this section, we will classify different scenarios by the relative ordering of these scales and discuss each case in detail. 

We split the scenarios into two categories depending on the $L_{\rm SN}$ and $\ell_*$ relation. Case I is the low-confusion regime where $L_{\rm SN}<\ell_*$, corresponding to the $N_{\rm eff}<1$, and we further discuss three subcases in this category depending on values of $\sigma_L$. Case II is the highly confused regime defined by $L_{\rm SN}>\ell_*$, corresponding to $N_{\rm eff}>1$.  

\begin{figure}[htbp!]
\begin{center}
\includegraphics[width=\linewidth]{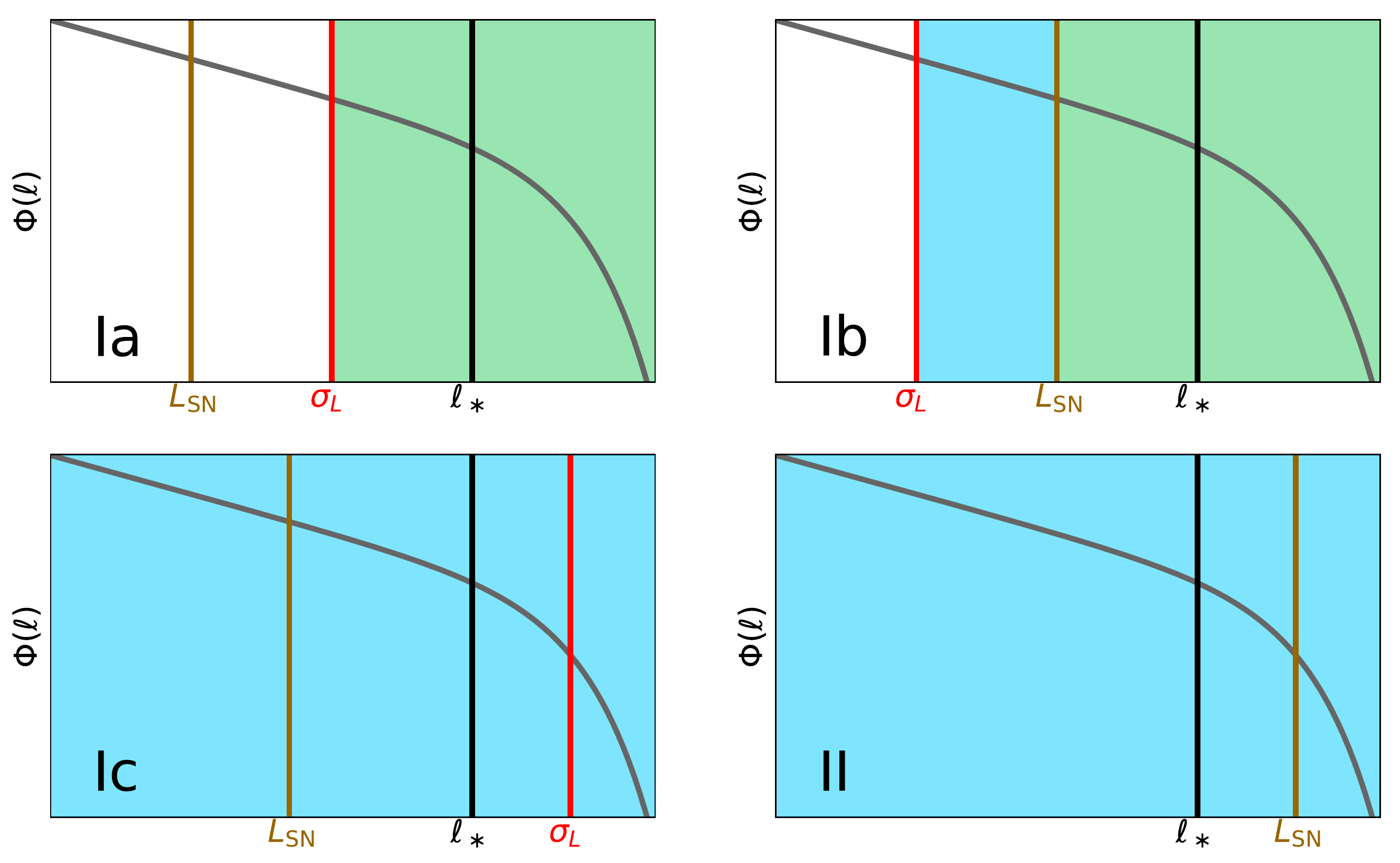}
\caption{\label{F:cases} Ordering of $\{L_{\rm SN}, \sigma_L, \ell_\ast\}$ in each case discussed in Sec.~\ref{S:Sch_noise}. Case I is defined by $L_{\rm SN}<\ell_*$, corresponding to the $N_{\rm eff}<1$ low-confusion regime, and its three subcases in this category as determined by the position of $\sigma_L$. Whereas the Case II is the highly confused regime defined by $L_{\rm SN}>\ell_*$, corresponding to $N_{\rm eff}>1$. The blue shaded regions are where IM is the optimal strategy, and the green shaded regions mark the scales above the optimal threshold when the GD observable is the optimal strategy.}
\end{center}
\end{figure}

Fig.~\ref{F:cases} summarizes the schematic ordering of these categories, and the shaded regions mark the optimal observing strategy for each case discussed below.

\subsubsection{Case I: $L_{\rm SN}<\ell_\ast$}
Here we have a relatively low number density, with 
$L_{\rm SN}<\ell_\ast$, approximately corresponding to the $N_{\rm eff} < 1$ regime.
We will thus apply the $P(D)$ calculation to derive the $P(L)$ and the optimal observable. 

\begin{figure*}[htbp!]
\begin{center}
\includegraphics[width=\linewidth]{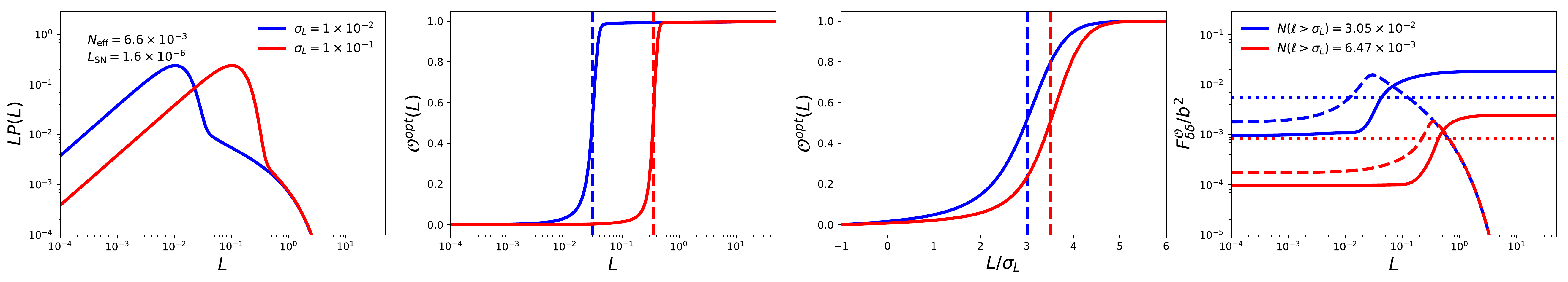}
\caption{\label{F:Ia}Two examples of case Ia. \textbf{Left:} $P(L)$ after convolving with $\sigma_L=0.01$ (blue) and $0.1$ (red). \textbf{Middle left:} optimal observables (solid lines). The dashed lines are the optimal threshold for the step function observable, i.e. the peak of the dashed curve in the right panel. \textbf{Middle right:} same as the middle left panel, but with $L/\sigma_L$ on the x-axis on a linear scale. \textbf{Right}: The integrated Fisher information for the optimal observable (solid), Fisher information of the step function observable as a function of step position (dashed), and the Fisher information of the linear observable (dotted). }
\end{center}
\end{figure*}

\paragraph{\rm \textbf{Case Ia}: $\mathbf{L_{\rm SN}<\sigma_L< \ell_\ast}$}
We first consider the case of intermediate instrument noise, i.e., between $L_{\rm SN}$ and $\ell_\ast$. Fig.~\ref{F:Ia} shows two examples in this case with different $\sigma_L$. This is the regime where GD works well: the instrument noise is much smaller than $\ell_\ast$, and the voxels with $L\gtrsim \sigma_L$ do not suffer from confusion noise. Therefore, as expected, the optimal observable here is close to a step function with a transition at a few times $\sigma_L$ (Fig.~\ref{F:Ia}, two middle panels). The optimal step function has a threshold at $\sim 3 \sigma_L$ (dashed vertical lines in the two middle panels), and this optimal step function observable indeed captures nearly the optimal information, as shown in the right panel of Fig.~\ref{F:Ia}. This indicates that GD using a threshold at a few $\sigma$ is the optimal strategy. 

We also note from the solid curves in the right panel of Figure \ref{F:Ia} that the information content is dominated by voxels with total luminosity within an order of magnitude of the optimal threshold value at $\sim 3 \sigma_L$.

The total optimal Fisher information $F^{\rm  opt}_{\delta\delta}/b^2$ in this case should be of the order $N (\ell > \sigma_L)$, the number of sources per voxel above $\sigma_L$, since we can count sources brighter than the noise level without confusion. This is consistent with the results in the right panel of Fig.~\ref{F:Ia}, though $F^{\rm  opt}_{\delta\delta}/b^2$ is slightly lower than $N (\ell > \sigma_L)$ owing to instrumental noise $\sigma_L$. 

\paragraph{\rm \textbf{Case Ib}: $\mathbf{\sigma_L< L_{\rm SN}< \ell_\ast}$}
\begin{figure*}[htbp!]
\begin{center}
\includegraphics[width=\linewidth]{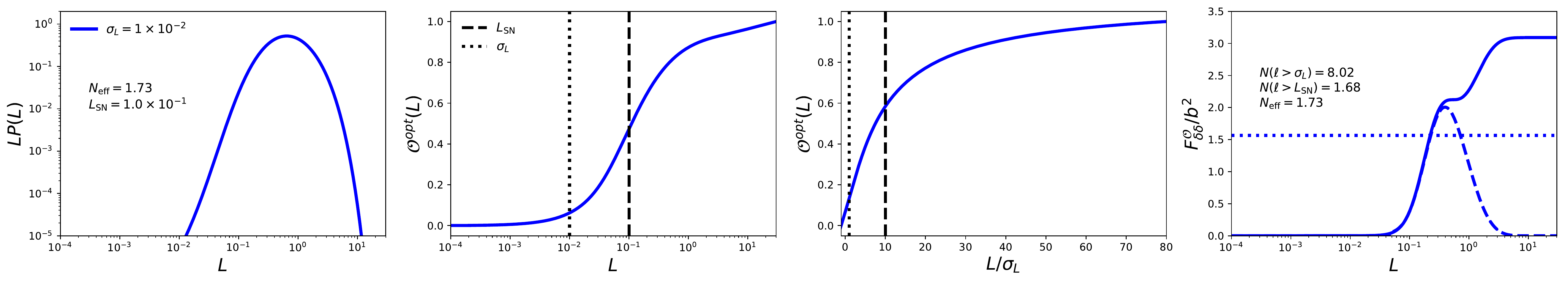}
\caption{\label{F:Ib} Case Ib example. 
\textbf{Left:} $P(L)$ after convolving with $\sigma_L=0.01$. \textbf{Middle left:} optimal observables (solid line). The dashed and dotted lines are $L_{\rm SN}$ and $\sigma_L$, respectively. \textbf{Middle right:} same as the middle left panel but plot with $L/\sigma_L$ in the x-axis. \textbf{Right}: integrated Fisher information for the optimal observable (solid), Fisher information of the step function observable as a function of step position (dashed), and the Fisher information of the linear observable (dotted).}
\end{center}
\end{figure*}

We now consider the low-noise regime, $\sigma_L < L_{\rm SN}$. Here the optimal observable is an intermediate between the IM and GD observables. Fig.~\ref{F:Ib} shows one scenario in this regime.
As in case Ia, one might naively apply a GD threshold at a few times $\sigma_L$. In the case Ia scenario, the voxel fluxes above the threshold are indeed ``detected'' since they rise above the instrumental noise and confusion. However, in case Ib, voxels above this threshold typically contain multiple sources with $\ell$ above the threshold, and the confusion noise from sources below the threshold is larger than the the sources at or just above the threshold. The regime of voxel fluxes $\sigma_L < L < L_{\rm SN}$ is thus more amenable to the IM technique. Individual sources can be detected with a threshold $L_{\rm th} \gtrsim L_{\rm SN}$ because only those sources rise above the confusion noise. 
 
 The resulting optimal observable can thus be understood as a hybrid between the two methods, detecting individual sources in the brightest voxels ($L > L_{\rm SN}$), and benefiting from IM in the fainter voxels that still rise above the instrumental noise ($\sigma_L < L < L_{\rm SN}$).

Fig.~\ref{F:Ib} indeed shows that neither the pure IM (linear) nor the pure GD (step function) observables capture the optimal information. The Fisher information for the optimal observable gains information in two stages, corresponding to the IM and GD parts, respectively. The total optimal Fisher information falls between $N(\ell>\sigma_L)$ and $N(\ell>L_{\rm SN})$, captured by GD and IM observables, respectively.

The detailed shape of the optimal observable depends on the luminosity function. In practice, we usually do not have sufficient knowledge of the source luminosity function, and it might be difficult to derive the optimal observable within our formalism. From our analysis, we know that the optimal observable in case Ib is GD above a threshold around $L_{\rm SN}$ and IM between that and another threshold around $\sigma_L$. Therefore, in practice, the optimal observable in the case Ib regime could be designed by choosing these two threshold scales, and by considering a linear function in between and a constant plateau about the upper threshold. By trying a range of values for both thresholds, the optimal threshold can be determined as the one giving the minimum shot-noise level in the power spectrum.

\paragraph{\rm \textbf{Case Ic}: $\mathbf{L_{\rm SN}< \ell_\ast<\sigma_L}$}
\begin{figure*}[htbp!]
\begin{center}
\includegraphics[width=\linewidth]{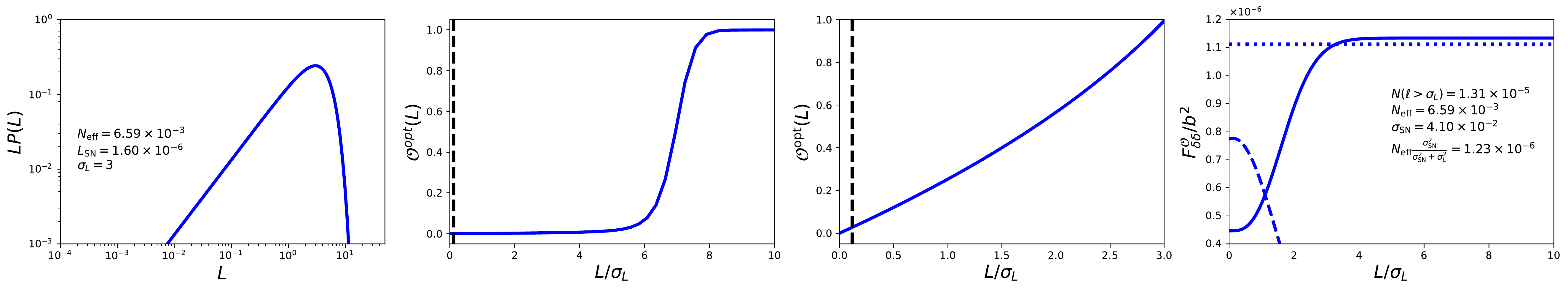}
\caption{\label{F:Ic}Case Ic example. \textbf{Left:} $P(L)$ after convolving with $\sigma_L=3$. \textbf{Middle Left:} optimal observables (solid line). The dashed line is the optimal threshold for the step function observable, i.e. the peak of the dashed curve in the right panel. \textbf{Middle right:} optimal observable zoomed in to around $\sigma_L=1.5$. \textbf{Right}: The integrated Fisher information for the optimal observable (solid), Fisher information of the step function observable as a function of step position (dashed), and the Fisher information of the linear observable (dotted).}
\end{center}
\end{figure*}

The final scenario in the $L_{\rm SN} < \ell_\ast$ ($N_{\rm eff} < 1$) regime is that of a very large instrument noise, $\sigma_L > \ell_\ast$. This is the case of noisy surveys, where only sources in the bright exponential tail of the Schecter function rise above the instrument noise.

Fig.~\ref{F:Ic} shows an example of case Ic. At first sight, the middle left panel appears to suggest that
the optimal observable is close to a GD step function with a threshold at $\sim 6 \sigma_L$.
However, when we consider the actual step function, we see first that the optimal threshold lies at $\sim 1 \sigma_L$,
and second (from the right panel) that its information content is far from optimal.
Inspecting the optimal observable in more detail, we see from the right panel that its information content is dominated by voxel luminosities up to $L \lesssim 3 \sigma_L$. In this regime, as shown by the middle right panel, the optimal observable is close to linear (and voxel luminosities are noisy). Thus, the optimal observable is closer to the IM observable. This interpretation is confirmed by considering in the right panel the information contained in the IM observable, which is indeed close to optimal.

Since sources brighter than the noise are not confused ($L_{\rm SN} < \sigma_L$), one might a priori expect GD to be the optimal strategy, just like in case Ia. The reason the present case is different is that sources brighter than the instrument noise are in the exponential tail of the Schechter function.
A detection threshold at a few times $\sigma_L$ that unambiguously distinguishes sources above the threshold from noise fluctuations would detect only a very small number of sources and throw away information in almost all voxels.
A slightly better approach is GD with a low threshold at $L \sim \sigma_L$. In this case, there are many false detections owing to the high instrumental noise, but a larger number of sources are probed. As discussed above, the approximately optimal approach is the IM observable, which gives an information content determined by the effective number of sources and the instrument noise suppression, $F^{\rm IM}_{\delta \delta}/b^2 = N_{\rm eff} \, \sigma^2_{\rm SN}/(\sigma^2_{\rm SN} + \sigma^2_L)$, larger than the information content given by the number of objects that can be detected, ($F^{\rm GD}_{\delta \delta}/b^2)\sim N(\ell > \sigma_L) \ll N_{\rm eff}$. 

\subsubsection{Case II: $\ell_\ast<L_{\rm SN}$}

\begin{figure}[htbp!]
\begin{center}
\includegraphics[width=\linewidth]{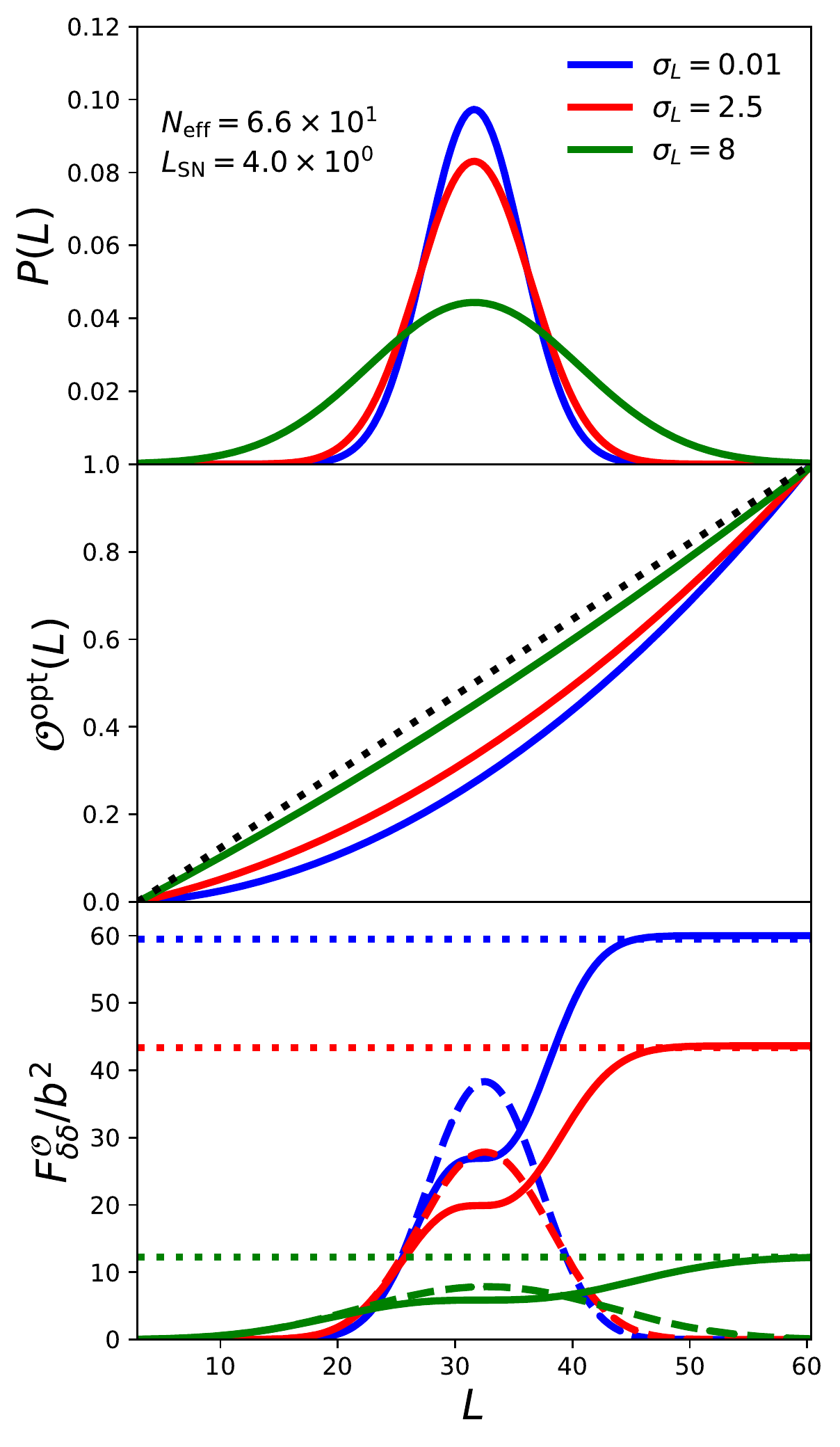}
\caption{\label{F:II}Case II with three different $\sigma_L$ levels. \textbf{Top:} $P(L)$ after convolving with $\sigma_L$. \textbf{Middle:} optimal observables (solid lines). The dashed line is the linear observable for reference. \textbf{Bottom}: The integrated Fisher information for the optimal observable (solid), Fisher information of the step function observable as a function of step position (dashed), and the Fisher information of the linear observable (dotted).}
\end{center}
\end{figure}

The defining criterion of case II, $\ell_* < L_{\rm SN}$, approximately corresponds to a large effective number of sources per voxel, $N_{\rm eff} > 1$. The $P(L)$ function here (at least in the $N_{\rm eff} \gg 1$ limit) can be approximated by a Gaussian with mean $\mu$ and variance $\bar{\sigma}^2$ given by
\begin{equation}
\mu=\int d\ell\, V_{\rm vox}\, \Phi(\ell)\,\ell=
\langle \hat{L}\rangle,
\end{equation}
and 
\begin{equation}
\bar{\sigma}^2=\int d\ell\,V_{\rm vox} \,\Phi(\ell)\,\ell^2+\sigma_L^2=\sigma_{\rm SN}^2+\sigma_L^2.
\end{equation}

Fig.~\ref{F:II} shows results for three different noise levels, corresponding to the three subclasses of case II: $\sigma_L<\ell_\ast<L_{\rm SN}$ (blue), $\ell_\ast<\sigma_L<L_{\rm SN}$ (red), and $\ell_\ast<L_{\rm SN}<\sigma_L$ (green). 

As in the $N\gg 1$ case in the toy model (Sec.~\ref{S:toyNl}), we derive the optimal observable to be the sum of a linear and a quadratic term,
\begin{equation}
\mathcal{O}^{\rm opt}(L)=L'+\frac{\langle \hat{L}^2\rangle}{2\,\mu\,\bar{\sigma}^2}L'^2,
\end{equation}
where $L'=L-\mu$. The quadratic term has a negligible contribution to the optimal Fisher information, similarly to the toy model, so IM (the linear function observable) is the optimal strategy, and the optimal Fisher information $F^{\rm opt}_{\delta \delta} \sim F^{\rm IM}_{\delta \delta}/b^2$ has the upper bound $N_{\rm eff}$ (see Eq.~\ref{E:F_IM_Sch}), and drops as the noise goes up.

\subsection{Schechter Luminosity Function Model Summary}
In this section, we explored four different scenarios defined by different ordering of $\ell_*$, $L_{\rm SN}$, and $\sigma_L$. Our formalism is not restricted to the IM or GD observable, but we found that in most cases either IM or GD is indeed the optimal strategy for mapping LSSs. Only in case Ib will an alternative strategy defined as the hybrid of the two will outperform a pure IM or pure GD observable, but case Ib is a very rare situation. None of the future surveys discussed in Sec.~\ref{S:surveys} are in the case Ib regime. Therefore, we conclude that the GD / IM dichotomy captures most of the optimal strategy in reality.

\section{Optimal Strategy for IM Experiments}\label{S:surveys}
We now apply the formalism we have developed to proposed and ongoing IM experiments. By simply calculating $L_{\rm SN}$, $\ell_*$, and $\sigma_L$ from experimental parameters and empirical line luminosity functions, we can categorize a survey into one of the cases in Sec.~\ref{S:Sch_noise}, and identify its optimal observable. 

As discussed in Sec.~\ref{S:NeffLsn}, there exist ambiguous regimes where the cases will be classified as case I ($L_{\rm SN}<\ell_*$), but the confusion is significant ($N_{\rm eff}>1$). Therefore, we also calculate $N_{\rm eff}$ for each experiment, and we label these cases I/II as they are intermediate, instead of classifying them into either one of the cases.

Below we consider several experiments targeting different spectral lines across redshift. The results for all the surveys and lines we discuss below are summarized in Table~\ref{T:surveys}. We present the relevant parameters of each survey and leave the details in Appendix~\ref{A:units}.

An important potential caveat to the discussion here is that we only include the instrumental noise as the noise term $\sigma_L$. In reality, astrophysical foreground contaminations, for example, are another source of noise, and their fluctuations could be much higher than the instrumental noise without any foreground mitigation procedure. These foregrounds may include both local contributions from the Milky Way galaxy and emissions from extragalactic sources. Fortunately, these foregrounds are in principle distinguishable from the line signal of interest because of their distinct spectral and spatial signatures, often being much smoother spectrally than the signal that enables us to remove them with the strategies advocated for foreground cleaning in 21 cm IM measurements \citep{2011PhRvD..83j3006L, 2012ApJ...756..165P,2015ApJ...815...51S}. Quantifying the effect of residual foregrounds requires a more sophisticated model, which is outside the scope of this work.

\subsection{SPHEREx}

SPHEREx is a planned space mission for an all-sky near-infrared spectro-imaging survey \citep[ \url{http://spherex.caltech.edu}]{2014arXiv1412.4872D}. SPHEREx would carry out the first all-sky spectral survey at wavelengths between 0.75 and 2.42 $\mu$m (with spectral resolution $R=41$), between 2.42 and 3.82 $\mu$m (with $R=35$), between 3.82 and 4.42 $\mu$m (with $R=110$), and between 4.42 and 5.00 $\mu$m (with $R=130$), with a pixel size of $6.2''$. We take the $5\sigma$ sensitivity to be $m_{AB}=19.5$ and 22 per spectral channel, which is approximately the expected sensitivity in the all-sky and the deep regions (2 $\times \sim 100$ deg$^2$), respectively. SPHEREx is able to detect multiple lines, including H$\alpha$, H$\beta$, [OIII], and Ly$\alpha$, at different redshifts. Here we discuss the cases of H$\alpha$ and Ly$\alpha$.

\paragraph{\rm H$\alpha$} SPHEREx can detect the H$\alpha$ line at $0.1<z<5$. We adopt the {H$\alpha$} luminosity function at $z=2.23$ from \citet{2013MNRAS.428.1128S}: a Schechter function with $\rm log_{10}\,\phi^\ast=-2.78\ Mpc^{-3}, \ log_{10}\,\ell_*=42.87\ \rm{erg/s},\ and \ \alpha=-1.59$. We then derive from the luminosity function and instrument parameters that $L_{\rm SN}/\ell_*=5.8\times 10^{-5}$, $N_{\rm eff}=2.2\times 10^{-2}$, and $\sigma_L/\ell_*=0.19$ (deep regions) and $\sigma_L/\ell_*=1.9$ (all-sky). The all-sky survey is clearly in the case Ic regime, where IM is optimal. As for the deep regions, at first sight, it is in the case Ia regime ($L_{\rm SN}<\sigma_L<L_\ast$), where GD is the optimal strategy. However, since $\sigma_L$ is close to $\ell_*$, we are really at the boundary between the case Ia and the case Ic scenario, the latter suggesting that IM is preferred. Since we are in this gray area between the two regimes, an explicit calculation is required to check which approach is optimal. We thus computed the Fisher information for the linear and step function observables and found that the two approaches have the similar performance. Therefore, we label it with IM/GD as there is no preferred approach in this case.

\paragraph{\rm Ly$\alpha$}The Ly$\alpha$ line from high redshifts ($5.2<z<8$) also falls within the SPHEREx bands. Here we use the Ly$\alpha$ luminosity function at $z=5.56$ from \citet{2011A&A...525A.143C}: a Schechter function with $\rm \phi^*=9.2\times 10^{-4}\ Mpc^{-3}, \ log_{10}\,\ell_*=42.72\ \rm{erg/s},\ \alpha=-1.69$, and from this we get $L_{\rm SN}/\ell_*=2.1\times 10^{-4}$, $N_{\rm eff}=3.2\times 10^{-2}$, and $\sigma_L/\ell_*=6.4$ (deep regions) and $\sigma_L/\ell_*=64$ (all-sky). Both are in the case Ic regime, so IM is again the optimal strategy.

\subsection{CDIM}
The Cosmic Dawn Intensity Mapper \citep[CDIM,][]{2016arXiv160205178C} is a NASA Probe Study designed for Cosmic Dawn and Epoch of Reionization studies, probing Ly$\alpha$, H$\alpha$ and other spectral lines through cosmic history as part of its science goals. It plans to cover the wavelength range of $\rm 0.75-7.5\, \mu m$, with a spectral resolution of $R=300$ and 1 arcsec$^2$ pixel size. The planned $\sim 30$ deg$^2$ deep surveys would reach a 5$\sigma$ point-source sensitivity of $\rm m_{AB} = 22.5$. We calculate the H$\alpha$ and Ly$\alpha$ line signals using the same luminosity functions described in the SPHEREx analysis above. 

\paragraph{\rm H$\alpha$} For H$\alpha$ at $z=2.23$, we found $L_{\rm SN}/\ell_*=1.8\times 10^{-9}$, $N_{\rm eff}=4.9\times 10^{-5}$, and $\sigma_L/\ell_*=9.8\times 10^{-3}$. This is clearly inside the case Ia regime ($L_{\rm SN}<\sigma_L<\ell_*$), where the sources above the instrumental noise can be detected without confusion, so GD is the optimal strategy and the Fisher information is $\sim N(>\sigma_L)$.

\paragraph{\rm Ly$\alpha$} For Ly$\alpha$ at $z=5.56$, we have $L_{\rm SN}/\ell_*=8.0\times 10^{-8}$, $N_{\rm eff}=1.4\times 10^{-4}$, and $\sigma_L/\ell_*=0.68$. This is at the boundary between the Ia and Ic scenarios, as with the SPHEREx H$\alpha$ (deep regions) case, where IM and GD observables have the similar performance, so we label it with IM/GD.

\vskip 7pt

We remind the reader that, to reach the conclusion that thresholded detection of individual lines is optimal for this survey, we have assumed that residual foregrounds can be ignored so that only the instrumental noise (and the shot noise in the line-emitting galaxies) enters the problem. Incorporating foregrounds (including continuum emission from extragalactic sources) in a realistic way may alter the conclusion on the optimal observable.

\subsection{HETDEX} 
The Hobby-Eberly Telescope Dark Energy Experiment \citep[HETDEX,][\url{ www.hetdex.org}]{2008ASPC..399..115H} is a wide-field survey covering 300 deg$^2$ at the north Galactic cap.
Its main science goal is to detect 0.8 million Ly$\alpha$-emitting (LAE) galaxies within $1.9<z<3.5$ to provide a direct probe of dark energy at $z\sim 3$. The survey will have a $\rm 3''\times 3''$ pixel size, and the spectral resolution is $R=800$. The quoted sensitivity for 1200 s exposures per field is approximately $\rm 6\times 10^{-17} erg/s/cm^2$ (5$\sigma$), so we set $\rm \sigma_L=1\times 10^{-17} erg/s/cm^2$ in our calculation.

\paragraph{\rm Ly$\alpha$} Here we consider the Ly$\alpha$ measurement at $z=2.5$ using the luminosity function from \citet{2011A&A...525A.143C} in their $1.95<z<3$ redshift bin (a Schechter function with $\rm \phi^\ast=7.1\times 10^{-4}\ Mpc^{-3}, \ log_{10}\,\ell_*=42.70\ \rm{erg/s},\ \alpha=-1.6$). Then, we derive $L_{\rm SN}/\ell_*=1.2\times 10^{-8}$, $N_{\rm eff}=1.3\times 10^{-4}$, and $\sigma_L/\ell_*=9.3\times 10^{-2}$, which is also the in the Ia regime, so that line detection is the optimal strategy. 

\vskip 7pt

Although our calculations for CDIM and HETDEX for detecting Ly$\alpha$ indicate that galaxy/line detection is a better option than IM, we have assumed that the Ly$\alpha$ emission comes from point sources. However, Ly$\alpha$ photons are very often rescattered with nearby neutral hydrogen before they escape from galaxies, and thus the Ly$\alpha$ emission is extended. According to radiative transfer simulations, the extended Ly$\alpha$ halos have a size of tens or even hundreds of kpc \citep{2005ApJ...628...61C,2007ApJ...657L..69L,2010ApJ...708.1048K,2011ApJ...739...62Z}, which is comparable to the pixel size we consider here (the comoving voxel dimension in our Ly$\alpha$ calculation is $8.4\times 0.027\times0.027\, Mpc/h$ and $3.5\times 0.059\times0.059\, Mpc/h$ for CDIM and HEDEX, respectively). As a result, it is possible that IM is a better way to capture the extended Ly$\alpha$ emission; a more detailed investigation is needed to quantify the best observable for the Ly$\alpha$ line. 

Another potential caveat is that the ``GD'' we discuss in this work is only based on the targeting line emission, while no external information is used for source detection. In reality, however, sources might be {\it detected} based on their full spectrum, and the line is then used to get its redshift. This is closer to the observing strategy for HETDEX. Since our model is not applicable for this type of survey strategy, a more sophisticated formalism is needed in order to quantify its ability to extract the LSS information. 

\subsection{TIME}
TIME is a grating spectrometer dedicated to probe the [CII] line at $5.3<z<8.5$ \citep{2014SPIE.9153E..1WC}. The instrument has a spectral resolution of $R=150$ and a pixel size of $0'.45$. The noise-equivalent intensity (NEI) is around $\rm 10^{6} - 10^{7} Jy \sqrt{sec}/sr$, and we adopt $\rm NEI=4\times 10^{6}Jy \sqrt{sec}/sr$ for the calculation. The proposed 1000 hr survey gives an integration time per pixel of $\rm t_{pix}=100\ hr$, leading to $\rm\sigma_L=NEI/\sqrt{2\,t_{pix}}=4.71\times 10^{3}\ Jy/sr$.

\paragraph{\rm [CII]} We now calculate the performance of TIME probing [CII] at $z=6$. For the luminosity function, we adopt the semianalytic model from \citet{2016MNRAS.461...93P} (a Schechter function with $\rm \phi^*=(\rm{ln 10})10^{-2.95} \ Mpc^{-3}, \ log_{10}\,\ell_*=7.80\ \rm{L_\odot},\ \alpha=-1.77$). From these we get $L_{\rm SN}/\ell_*=1.9\times 10^{-2}$, $N_{\rm eff}=0.75$,  and $\sigma_L/\ell_*=2.17$. This is in the case Ic regime, where IM is the optimal strategy.

\subsection{COMAP} 
The CO Mapping Array Pathfinder \citep[COMAP,][]{2016AAS...22742606C} aims at tracing star formation through cosmic time with the CO rotational transition lines. COMAP will observe in the  30-34 GHz window with a 40 MHz spectral resolution, corresponding to CO(1-0) at $2.4<z<2.8$ and CO(2-1) at $5.8<z<6.7$. Following the formalism and the instrument parameters of the Pathfinder in \citet{2016ApJ...817..169L}, we obtain a pixel size of $2'.55$ and a system noise of $23\ \mu K$. 
\paragraph{\rm CO(1-0)} We now consider the CO(1-0) line at $z=3$. For the luminosity function at $z=3$ , we take the averaged value of each of the three Schechter function parameters for $z=2$ and $z=4$ in \citet{2016MNRAS.461...93P}: $\phi^*=(\rm{ln 10})10^{-2.79} \ Mpc^{-3}, \ log_{10}\,\ell_*=7.28\ \rm{Jy\ km\ s^{-1}\ Mpc^2},\ \alpha=-1.62$. From these we get $L_{\rm SN}/\ell_*=1.4\times 10^{-1}$, $N_{\rm eff}=2.5$, and $\sigma_L/\ell_*=13$, so this is near the borderline of the Ic ($L_{\rm SN}<\ell_*<\sigma_L$) and II regimes ($N_{\rm eff}>1$), where IM is the optimal strategy in both cases.

\subsection{CHIME}
The Canadian Hydrogen Intensity Mapping Experiment \citep[CHIME,][]{2014SPIE.9145E..22B} is a cylindrical interferometer designed to measure the neutral hydrogen HI power spectrum at $0.8<z<2.5$. We consider the HI signal at $z = 1$. The instrument has a $15-25$ arcmin angular resolution, and we adopt $15$ arcmin as the pixel size. The frequency resolution is 390 kHz \citep{2014SPIE.9145E..22B}, and the noise level at $z=1$ is $\sigma_T=2.9\times 10^{-4}$K for 1.4 yr of integration, calculated from the survey parameters given in \citet{2014SPIE.9145E..22B} (see Appendix.~\ref{A:units} for the derivation). 

For the HI luminosity function, we use the local ($z<0.06$) HI observations from \citet{2010ApJ...723.1359M}, in which the HI mass function is fitted with a Schechter function with $\phi_\ast=4.8\ \rm h^3_{70}\ Mpc^{-3}\ dex^{-1}, log(M_\ast/M_\odot)+2log\ h_{70}=9.96$, and $\alpha=-1.33$, and we ignore redshift evolution from $z=1$ to the present day.
See Appendix~\ref{A:units} for converting the HI mass function to the luminosity function.

With this information in hand, we get $L_{\rm SN}/\ell_\ast=0.63$, $N_{\rm eff}=4.2$, and $\sigma_L/\ell_\ast=3.4$, which is again near the borderline of Ic and II regimes, where IM is optimal for both cases. We stress again that this is a calculation for an idealized situation that ignores foreground effects. 

\vskip 7pt

\begin{table*}[!ht]
\centering
\caption{\label{T:surveys}Summary of the survey targets and their expected $\sigma_L$, $\ell_\ast$, and $L_{\rm SN}$ relation.}
\begin{tabular}{lllllllll}
\hline
\hline
 survey & Line& redshift & $\sigma_L/\ell_\ast$ & $L_{\rm SN}/\ell_*$ & Ls' Relation & $N_{\rm eff}$ & Case & Optimal Strategy \\
\hline
SPHEREx (deep regions) & H$\alpha$  & 2.23     & $0.19$                & $5.8\times 10 ^{-5}$ & $L_{\rm SN}<\sigma_L\lesssim \ell_*$ & $2.2\times 10^{-2}$ & Ia/Ic  & GD/IM\footnote{\label{N:IaIc}These cases are at the boundary of Ia and Ic, so we confirm that the IM is better than GD by numerically calculating their $P(L)$ and their Fisher information of the GD, IM, and optimal observable.}      \\
        & Ly$\alpha$ & 5.56     & $6.4$                & $2.1\times 10 ^{-4}$ & $L_{\rm SN}<\ell_*<\sigma_L$         & $3.2\times 10^{-2}$ & Ic     & IM       \\
SPHEREx (all-sky) & H$\alpha$  & 2.23     & $1.9$                & $5.8\times 10 ^{-5}$ & $L_{\rm SN}<\ell_*<\sigma_L$ & $2.2\times 10^{-2}$ & Ic  & IM \\
        & Ly$\alpha$ & 5.56     & $64$                & $2.1\times 10 ^{-4}$ & $L_{\rm SN}<\ell_*<\sigma_L$         & $3.2\times 10^{-2}$ & Ic     & IM       \\
CDIM    & H$\alpha$  & 2.23     & $9.8\times 10 ^{-3}$ & $1.8\times 10 ^{-9}$ & $L_{\rm SN}<\sigma_L<\ell_*$        & $4.9\times10^{-5}$  & Ia     & GD       \\
        & Ly$\alpha$ & 5.56     & $0.68$ & $8.0\times 10 ^{-8}$ & $L_{\rm SN}<\sigma_L\lesssim\ell_*$         & $1.4\times 10^{-4}$ & Ia/Ic     & GD/IM\textsuperscript{\ref{N:IaIc}}      \\
HETDEX  & Ly$\alpha$ & 2.5      & $9.3\times 10 ^{-2}$ & $1.2\times 10 ^{-8}$ & $L_{\rm SN}<\sigma_L<\ell_*$         & $1.3\times10^{-4}$ & Ia     & GD       \\
TIME    & {[}CII{]}  & 6        & $2.17$               & $1.9\times 10 ^{-2}$ & $L_{\rm SN}<\ell_*<\sigma_L$ & $7.5\times10^{-1}$ & Ic  & IM      \\
COMAP   & CO(1-0)    & 3        & $13$                 & $1.4\times 10^{-1}$               & $L_{\rm SN}<\ell_*<\sigma_L$         & $2.5$ & Ic/II     & IM      \\
CHIME   & HI    & 1        & $3.4$                 & 0.63               &  $L_{\rm SN}<\ell_*<\sigma_L$         & $4.2$ & Ic/II     & IM      \\
\hline
\end{tabular}
\end{table*}

\vskip 7pt

The above analysis focuses on the 3D line IM experiments. Two-dimensional continuum surveys such as the cosmic infrared background (CIB) experiments are also worth discussing in this context, given that they usually suffer from confusion \citep{2013ApJ...779...32V,2017ApJ...850...37W,2017A&A...607A..89B}, which induces errors in measuring the properties of bright sources (e.g.  the position and flux error from confusion noise described in \citet{2001AJ....121.1207H}). Another common issue in the CIB experiments is the correlated confusion noise, which refers to the fact that the fluctuations from the faint, unresolved sources are spatially correlated with the bright sources. Our $P(D)$ formalism intrinsically captures the dependency of the density of all the sources and their underlying overdensity field $\delta$, regardless of the detection limit, and thus it is a suitable way to quantify the confusion in CIB. 
However, according to the observations, the CIB source luminosity function is close to a simple power law without an exponential cutoff at the bright end \citep{2013ApJ...779...32V}. Therefore unlike the Schechter function, there is no characteristic $\ell_*$ we can use to compare with $\sigma_L$ and $L_{\rm SN}$ to classify the regimes. A detailed $P(D)$ analysis is needed to study this different kind of luminosity function, and we leave it to the future works.

\section{Example Application: Pixel Size Optimization}\label{S:OptPix}

In this section, we use our framework to calculate the information content as a function of pixel (or beam) size. The choice of pixel size in a survey is a trade-off between confusion and instrumental noise, which are quantified by $L_{\rm SN}$ (or $N_{\rm eff}$) and $\sigma_L$, respectively. A smaller pixel size gives less confusion, but the instrumental noise $\sigma_L/\ell_*$ also changes according to the properties of the dominant noise source and how the integration time and collecting area scaled with the pixel size. The two effects cannot be treated independently if our observable is not a linear function, and thus it requires a full $P(D)$ analysis to construct the $P(L)$ distribution and then to derive the Fisher information.  

We consider changing the pixel size from $\Omega_{\rm pix}$ to $a \, \Omega_{\rm pix}$, while fixing the spectral bandwidth per voxel. Here $a$ is a rescaling parameter that quantifies the change in pixel size relative to a fiducial survey configuration, and we would thus like to compute $N_{\rm eff}$, $\sigma_L$, and ultimately the Fisher information in the new pixel, as a function of $a$. The voxel volume and $N_{\rm eff}$ trivially scale linearly with $a$. The exact effect on the instrumental noise per voxel depends on the details of the experiment and on how its specifications are varied as the pixel size is changed, as we will discuss in more detail below. With the variation in voxel size and $\sigma_L$, we can calculate the Fisher information in the new $a \, \Omega_{\rm pix}$ voxel. However, it is not sufficient to simply consider the variation (with $a$) in the Fisher information per voxel. A smaller pixel size gives a larger number of pixels to constrain the underlying $\delta$ for a fixed survey region. Therefore, the meaningful quantity for the performance of different voxel size is $F(a)/a$, where $F(a)$ is the Fisher information of a single voxel with size $a \, \Omega_{\rm pix}$. The quantity $F(a)/a$ gives the information content on $\delta$ for a fixed survey region.

The scaling of $\sigma_L/\ell_\ast$ is derived from comparing the number of photons from a $\ell_\ast$ source and the rms of the number of photons from noise for a given integration time.

The number of photons $N_{\rm src}$ from a $\ell_*$ source per voxel per integration $t_{\rm int}$ is given by
 \begin{equation}
 N_{\rm src}=\frac{\ell_\ast}{4\pi D_L(z)^2}A_{\rm coll}t_{\rm int}.
\end{equation}
We assume that the instrument's collecting area $A_{\rm coll}$, is fixed by the aperture size, and we assume a fixed total integration time/survey duration and a fixed total sky coverage for the survey. If we change the angular size of pixel from $\Omega_{\rm pix}$ to $a \, \Omega_{\rm pix}$ by moving the focal length of the telescope, while fixing the physical configuration of the detector (the physical pixel size and number of pixels on the detector stay the same), the instantaneous field of view also scaled with $a$, and thus the integration time per pixel $t_{\rm int}$ becomes $a \, t_{\rm int}$ in order to preserve the total integration time of the survey.  Therefore, we get $N_{\rm src}\propto a$.

As for the noise, below we will focus on two simple scenarios for the instrumental noise scaling with pixel size: a read-noise-dominated case and a photon-noise-dominated case. We will apply these two scalings relative to a fiducial experiment given by the SPHEREx $H\alpha$ case, presented in Sec.~\ref{S:surveys}.

\paragraph{\it Photon-noise-dominated scenario}
For the photon noise, we assume that the dominant photon source from the sky is a uniform bright foreground, e.g. the zodiacal light in the optical/near infrared. Say this foreground has surface brightness $I$, which has units Jy sr$^{-1}$. The number of photons $N_I$ from $I$ per voxel per integration is thus
\begin{equation}
N_I=I\, A_{\rm coll}\,\Omega_{\rm pix}\,\delta\nu\,t_{\rm int} \propto a^2,
\end{equation}
where $\delta\nu$ is the bandwidth, and we take it unchanged while varying the pixel size. The photon noise is the Poisson noise of $N_I$, and thus the rms of photon noise $\sigma_{\rm ph}$ is 
\begin{equation}
\sigma_{\rm ph}=\sqrt{N_I} \propto a.
\end{equation}
Therefore, the scaling of $\sigma_L/\ell_*$ with $a$ is proportional to $\sigma_{\rm ph}/N_{\rm src}$, which is a constant independent of voxel size.
 
\paragraph{\it Read-noise-dominated scenario} 
For the read noise, that assuming we only read at the beginning and the end of the integration, and each read has rms $\sigma_{\rm read}$ electrons, the expected rms number of photon of read noise $\sigma_{\rm RN}$ thus does not scale with $a$. As a result, $\sigma_L/\ell_*=\sigma_{\rm RN}/N_{\rm src}$ scales with $1/a$.

\vskip 7pt

\begin{figure}[htbp!]
\begin{center}
\includegraphics[width=\linewidth]{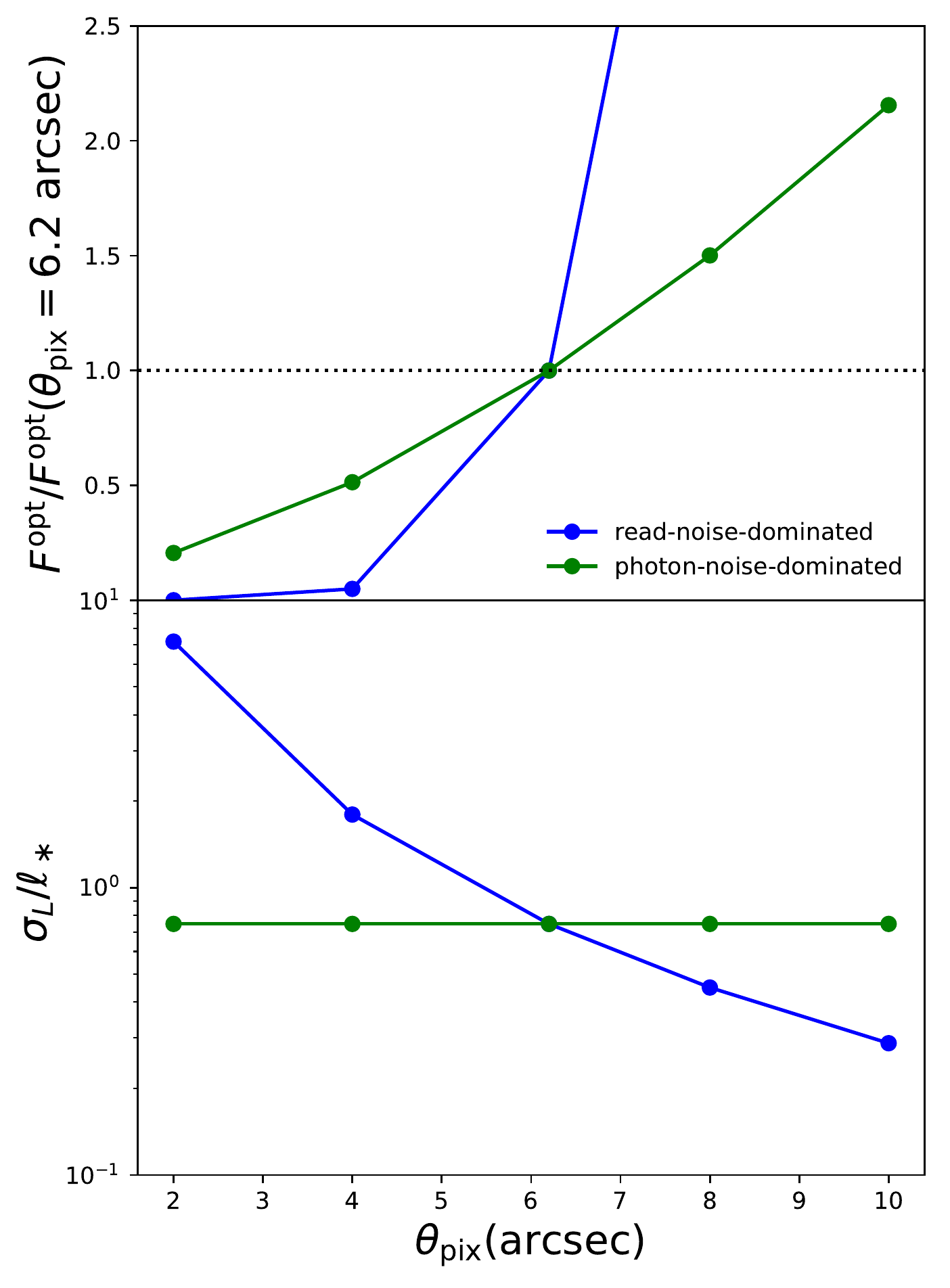}
\caption{\label{F:optpix} \textbf{Top}: The Fisher information of SPHEREx H$\alpha$ case with different pixel size. The Fisher information is normalized by the Fiducial 6.2 arcsec pixel size case. \textbf{Bottom}:  $\sigma_L/\ell_\ast$ ratio in each scenarios.}
\end{center}
\end{figure}

Fig.~\ref{F:optpix} shows the Fisher information ($F(a)/a$) for varying pixel/voxel size in the SPHEREx H$\alpha$ case, normalized by the Fisher information for the fiducial 6.2 arcsec pixel size. As shown in the plot, if the noise is dominated by read noise, increasing the voxel size will have a dramatic improvement on information gain, since this crosses the transition from Ic (IM) to Ia (GD) (see the bottom panel), and we expect a lot more information gain from individual detection. 

Here we only demonstrate a simple and idealized example of using this framework to quantify the information with different pixel sizes. We remind the reader that the scaling relation with pixel size we adopted here is not a unique behavior in the photon-noise- and read-noise-dominated cases. In reality, the pixel size can be changed in different ways (e.g. change the physical configuration of the pixels on the detector itself) and results in different scaling relation.

In addition, the discussion above assumes the fixed total survey volume. In reality, we can optimize the experiments by varying the survey volume as well. There is another trade-off between the survey volume and the depth ($\sigma_L$ in our context) for the given observing time. Increasing the total survey volume reduces the cosmic variance in the power spectrum. In this work, our formalism only accounts for the variance on the voxel-by-voxel basis, which corresponds to the shot noise in the power spectrum. In reality, cosmic variance is another noise source in the power spectrum that plays a significant role in the large-scale (low-k) mode uncertainty. To optimize the survey for probing the large-scale power spectrum, an analysis taking into account both the shot noise and cosmic variance is needed. We leave the consideration to future works.

\section{Conclusion}\label{S:conclusion}

We use a general ``observable'' as a weight function to turn the observed voxel flux map into the observable map that traces the LSS. The two well-studied approaches, GD and IM, are two special observable cases. The performance of observables is quantified by the Fisher information, and from it we derive the optimal observable, which is able to extract the full information content in the data.

We first work on a toy model assuming that all the targeting sources have the same flux $\ell$. By considering a range of source density $N$ (number of sources per voxel) and instrument noise level $\sigma_L$, we derive the optimal observable and its Fisher information for each case and compare it with the Fisher information of the GD and IM observables. In the toy model, we found that IM is preferred when the sources are either confused ($N>1$) or suppressed by the noise ($\sigma_L>\ell$). 

Next we move on to a more general model with the source population follows Schechter function form. Then, we identify four limiting regimes depending on the relative value of the three scales: $\{L_{\rm SN}, \sigma_L, \ell_*\}$. Again, we found that in the high-noise ($\sigma_L>\ell_*$, case Ic) or high-confusion ($N_{\rm eff}>1$ or $L_{\rm SN}>\ell_*$, case II) regime, the IM observable is preferred, as it reaches the performance of the optimal observable. Whereas on the opposite situation ($N_{\rm eff}<1$ and $\sigma_L<\ell_*$), we can further identify two distinct scenarios. The first one is where $L_{\rm SN}<\sigma_L<\ell_*$ (case Ia), such that all the voxels above the noise are not confused, so the detection with a threshold around $\sigma_L$ is the preferred strategy. The other scenario is where $\sigma_L<L_{\rm SN}<\ell_*$ (case Ib). In this case, the optimal strategy is the hybrid of the IM and GD observables. The IM observable is suitable for the voxels above noise but highly confused ($\sigma_L<L<L_{\rm SN}$), whereas for voxels above $L_{\rm SN}$, the voxel flux is dominated by a single bright source, and thus the GD is the favored choice for them.

Finally, we demonstrate the usage of this formalism with two applications. The first application is to identify the optimal strategy for the proposed (and ongoing) IM experiments (e.g. SPHEREx, TIME, COMAP). The second application is to calculate the information content for different pixel sizes in a survey. Although we have made some simplified assumptions in these two demonstrations, the formalism we developed here can be easily applied to optimizing the experiment parameters of interest with their own specification of noise and confusion level.

\acknowledgments
We are grateful to helpful discussions with the Caltech ObsCos group and the participants in the workshop ``Cosmological Signals from Cosmic Dawn to the Present''  held in the Aspen Center for Physics. Part of the research described in this paper was carried out at the Jet Propulsion Laboratory, California Institute of Technology, under a contract with the National Aeronautics and Space Administration. R.d.P. and O.D. acknowledge the generous support from the Heising-Simons Foundation.

\begin{appendices}

\def\fopt{$F^{\rm opt}_{\delta\delta}$}
\section{Proving \protect\fopt{}  $=\,F_{\delta\delta}$}\label{A:Fopt_proof}
Here we prove that the Fisher information per voxel of optimal observable $F^{\rm opt}_{\delta\delta}$ is equal to $F_{\delta\delta}$, the maximum Fisher information per voxel that any observable can possibly attain. Writing out each element in Eq.~\ref{E:Fisher} explicitly, we get
\begin{equation}
\begin{split}
\langle \hat{\mathcal{O}}^{\rm opt}\rangle &=\int dL\ P(L,\delta)\,\mathcal{O}^{\rm opt}(L)\\
&=\int dL\ P(L,\delta)\,\partial_{\delta} \ln P(L,\delta)\\
&=\partial_{\delta}\int dL\, P(L,\delta)=0\\
\end{split}
\end{equation}

\begin{equation}
\begin{split}
\langle (\hat{\mathcal{O}}^{\rm opt})^2  \rangle &=\int dL\ P(L,\delta)\,(\mathcal{O}^{\rm opt})^2(L)\\
&=\int dL\, P(L,\delta)\,(\partial_{\delta} \ln P(L,\delta))^2\\
&=\langle (\partial_{\delta}\ln P(L,\delta))^2 \rangle=F_{\delta\delta}
\end{split}
\end{equation}

\begin{equation}
\begin{split}
\partial_{\delta}\langle \hat{\mathcal{O}}^{\rm opt} \rangle &=\int dL\ \partial_{\delta}P(L,\delta)\,\mathcal{O}^{\rm opt}(L)\\
&=\int dL\,P(L,\delta)\,(\partial_{\delta}\,\ln P(L,\delta))\,\mathcal{O}^{\rm opt}(L)\\
&=\langle (\hat{\mathcal{O}}^{\rm opt})^2  \rangle=F_{\delta\delta},
\end{split}
\end{equation}
and thus
\begin{equation}
F^{\rm opt}_{\delta\delta}=\frac{\left(\partial_{\delta}\langle \hat{\mathcal{O}}^{\rm opt} \rangle\right)^2}
{\langle (\hat{\mathcal{O}}^{\rm opt})^2  \rangle-\langle \hat{\mathcal{O}}^{\rm opt}\rangle^2}=F_{\delta\delta}.
\end{equation}

\section{Comparing Linear and Quadratic Terms in the Toy Model $N\gg 1$ Optimal Observable}\label{A:lin_quad}
To explain why the quadratic term has a negligible contribution to the optimal Fisher information in the toy model $N\gg 1$ case (Sec.~\ref{S:toyNl}), below we explicitly calculate the components of Fisher information in Eq.~\ref{E:Fisher} for the linear ($\mathcal{O}^{\rm lin}(L)=L'$) and quadratic ($\mathcal{O}^{\rm quad}(L)\equiv \frac{\ell}{2\bar{\sigma}^2}L'^2$) terms in Eq.~\ref{E:RdP_Nl_oopt} respectively (note that $L'\equiv L-N\ell$, which is also the peak of the Gaussian $P(L)$ profile). The signals on these two components are
\begin{equation}
\begin{split}
&\partial_{\delta}\langle \hat{\mathcal{O}}^{\rm lin} \rangle=bN\ell\\
&\partial_{\delta}\langle \hat{\mathcal{O}}^{\rm quad} \rangle=\frac{bN\ell}{2}\left ( \frac{\ell^2}{\sigma_L^2+N\ell^2} \right ).
\end{split}
\end{equation}
Since this is in the $N\gg1$ regime, the signal from the quadratic term is always much smaller than from the linear term, regardless of the instrument noise $\sigma_L$.
The variance terms of the two observables are 
\begin{equation}
\begin{split}
&\langle (\hat{\mathcal{O}}^{\rm lin})^2 \rangle-\langle \hat{\mathcal{O}}^{\rm lin} \rangle^2=\bar{\sigma}^2-0=\sigma_L^2+N\ell^2\\
&\langle (\hat{\mathcal{O}}^{\rm quad})^2 \rangle-\langle \hat{\mathcal{O}}^{\rm quad} \rangle^2=\left ( \frac{\ell}{2\,\bar{\sigma}^2} \right )^2\left [ 3\,\bar{\sigma}^4-\left ( \bar{\sigma}^2 \right )^2 \right ]=\ell^2/2.
\end{split}
\end{equation}
Again, with the $N\gg 1$ condition, the contribution from the quadratic term is also negligible\footnote{To compare the Fisher information of purely linear observable with the full optimal observable (linear + quadratic), one also has to take into account the covariance term of these two observables $\langle \hat{\mathcal{O}}^{\rm lin}\hat{\mathcal{O}}^{\rm quad}\rangle$. Fortunately, this term vanished since it is an odd function with respect to the Gaussian $P(L)$ profile.}. Hence, the contribution of the quadratic term to the Fisher information is negligible, which implies a purely linear (IM) observable can reach the optimal performance.

\section{Explaining $F^{\rm IM}_{\delta\delta}\propto N$}\label{A:noiseless_lin}
The Fisher information of the IM observable is given by
\begin{equation}
F^{\rm IM}_{\delta\delta}=\frac{\left( \partial_{\delta}\langle \hat{L}\rangle\right)^2}
{\langle \hat{L}^2 \rangle-\langle \hat{L} \rangle^2},
\end{equation} 
where
\bea
\langle \hat{L}\rangle&=V_{\rm vox}\int d\ell\,\Phi(\ell)\,\ell \, \propto N\\
\langle \hat{L}^2\rangle&=V_{\rm vox}\int d\ell\,\Phi(\ell)\,\ell^2\, \propto N.
\eea
Below we will prove that the the numerator of $F^{\rm IM}_{\delta\delta}$ is proportional to $N^2$, and the denominator is proportional to $N$, and thus $F^{\rm IM}_{\delta\delta}$ is proportional to $N$.

The `signal' term is proportional to $N$ since $\partial_\delta\langle\hat{L}\rangle\,\propto\partial_\delta N=bN$. As for the variance $\sigma^2(\hat{L})=\langle \hat{L}^2 \rangle-\langle \hat{L} \rangle^2$,  we note the fact that we can divide each voxel into $N_{\rm sub}$ subvoxels, where the subvoxel fluxes $\hat{L}_i^{\rm sub}$ are independent of each other, so the total $\hat{L}$ is simply the sum of the subvoxel flux $\hat{L}_i^{\rm sub}$, and the variance $\sigma^2(\hat{L})$ is also the sum of the subvoxel variance $\sigma^2(\hat{L}_i^{\rm sub})$, $\sigma^2(\hat{L})=N_{\rm sub}\sigma^2(\hat{L}_i^{\rm sub})$, as the subvoxels are independent. The subvoxel variance is given by
\beq
\sigma^2(\hat{L}_i^{\rm sub})=\frac{V_{\rm vox}}{N_{\rm sub}}\int d\ell\,\Phi(\ell)\,\ell^2 -\left (\frac{V_{\rm vox}}{N_{\rm sub}}\int d\ell\,\Phi(\ell)\,\ell \right )^2.
\eeq
We have the freedom to choose $N_{\rm sub}$ large enough such that the second term is much smaller than the first term, so $\sigma^2(\hat{L}_i^{\rm sub})\propto\,V_{\rm vox}$ (and $N$), and the total voxel variance $\sigma^2(\hat{L})=N_{\rm sub}\sigma^2(\hat{L}_i^{\rm sub})$ is also proportional to $V_{\rm vox}$(and $N$). 

\def\lmin{$\ell_{\rm min}$}
\section{Different Choice of \protect\lmin{}}\label{A:Lmin}
\begin{figure}[htbp!]
\begin{center}
\includegraphics[width=\linewidth]{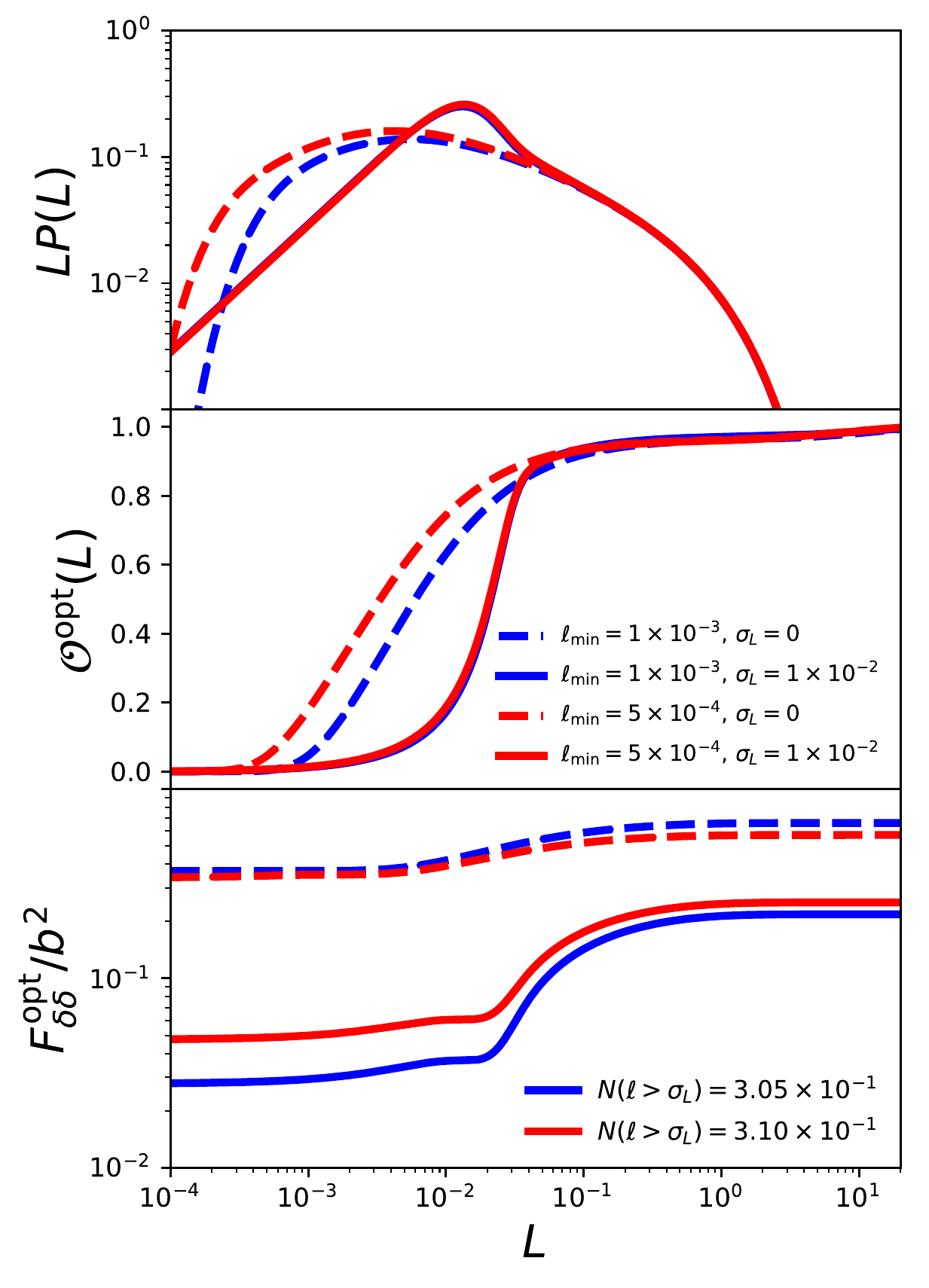}
\caption{\label{F:A_noise_Lmin}Fiducial Schechter function faint-end slope $\alpha=-1.5$ with and without instrumental noise $\sigma_L=0.01$ and using two different $\ell_{\rm min}$ . \textbf{Top:} $P(L)$ with (dashed lines) and without (solid lines) instrumental noise. \textbf{Middle:} optimal observables for each case. \textbf{Bottom:} The integrated Fisher information for the optimal observable.}
\end{center}
\end{figure}
Here we will justify that the choice of $\ell_{\rm min}$ does not affect the optimal observable and its information content. We compare the difference between fiducial $\ell_{\rm min}=10^{-3}$ and $\ell_{\rm min}=5\times10^{-4}$ cases, while keeping other parameters the same. The results are shown in Fig.~\ref{F:A_noise_Lmin}. The optimal observable is different in the absence of noise. However, if the instrumental noise is much higher than $\ell_{\rm min}$ (e.g. $\sigma_L=10^{-2}$ in this example), the effect of the artificial cutoff $\ell_{\rm min}$ is totally obscured by the noise, and thus both $\mathcal{O}^{\rm opt}$ and $F^{\rm opt}_{\delta\delta}$ are nearly identical in the two cases here. Therefore, we justify that the arbitrary choice of the $\ell_{\rm min}$ does not affect the optimal observable and Fisher information as long as the cutoff $\ell_{\rm min}$ is much lower than the instrument noise $\sigma_L$.

\section{Different Choice of $\alpha$}\label{A:alpha}
\begin{figure}[htbp!]
\begin{center}
\includegraphics[width=\linewidth]{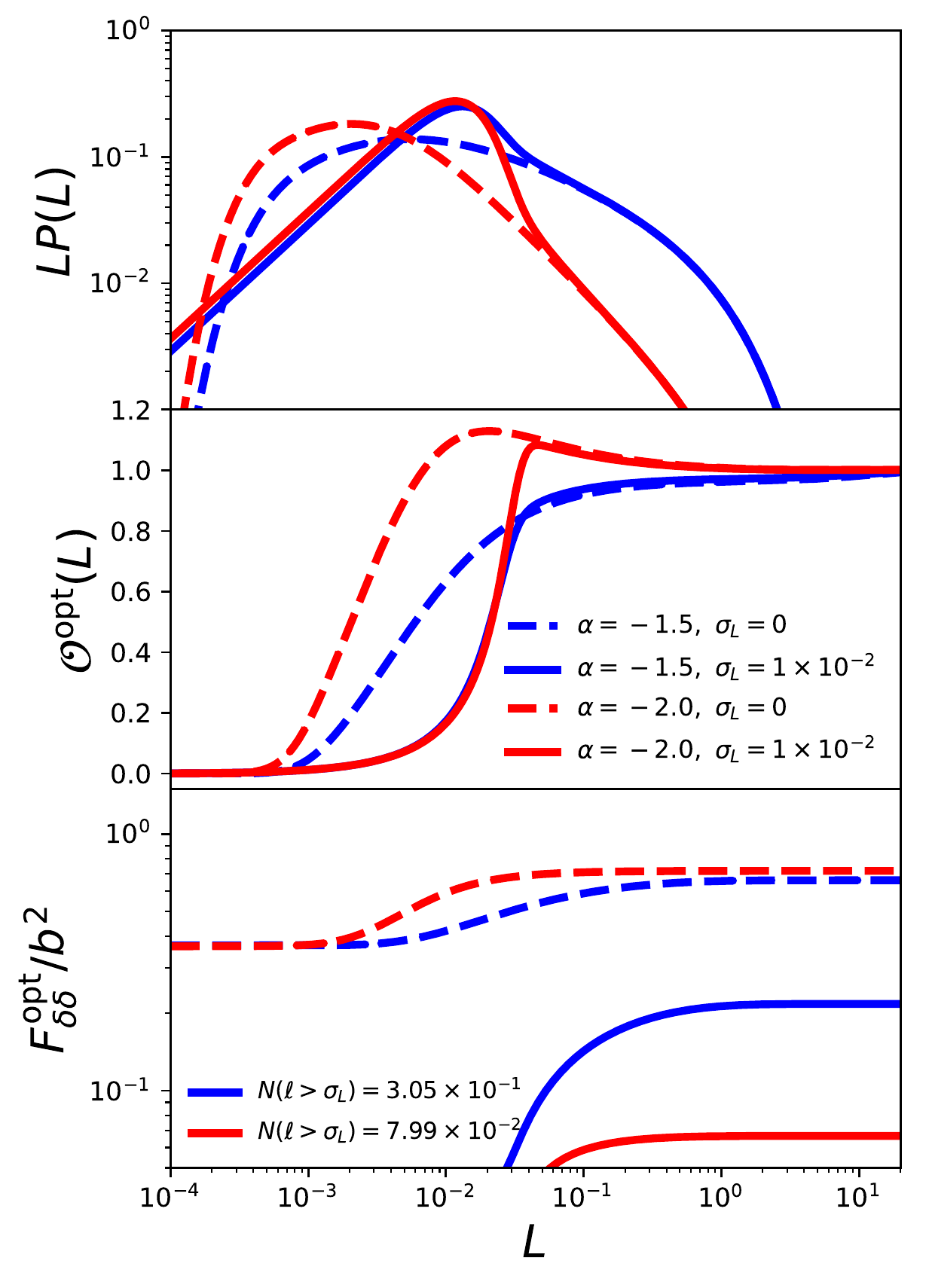}
\caption{\label{F:A_noise_alpha}Fiducial case with two different $\alpha$. \textbf{Top:} $P(L)$ of two different $\alpha$ with (solid lines) and without (dashed lines) instrumental noise noise $\sigma_L=0.01$. \textbf{Middle:} The optimal observables for each case. \textbf{Bottom:} The integrated Fisher information for the optimal observable.}
\end{center}
\end{figure}
Here we show how the different faint-end slope $\alpha$ affects the optimal observable and the Fisher information. Fig.~\ref{F:A_noise_alpha} compares the cases of fiducial $\alpha=-1.5$ with steeper faint-end slope $\alpha=-2$, while keeping other parameters the fiducial values. In the noiseless scenario, the optimal observable of the $\alpha=-2$ case has the step at lower $L$ compared to $\alpha=-1.5$ case. This naturally reflects the fact that there are more faint sources in the $\alpha=-2$ case. When a $\sigma_L=10^{-2}$ instrumental noise is applied, the difference is washed out by the noise. Another interesting feature is the peak in the $\mathcal{O}^{\rm opt}(L)$ function for the $\alpha=-2$ case, which can be explained by the fact that the voxels with luminosity around the peak are more likely to have multiple sources, whereas higher-$L$ voxels are mostly contributed by a single bright source. Because we assume a luminosity-independent bias, the source number density traces the underlying $\delta$ linearly, and thus the voxels around the peak are likely tracing the higher density field than the even brighter voxels. This does not happen in the $\alpha=-1.5$ case because of its lack of faint sources to reach this special regime.

\section{Unit Conversion of the Survey Parameters}\label{A:units}
In Sec.~\ref{S:surveys}, we derive the $\ell_\ast$, $L_{\rm SN}$, and $\sigma_L$ from the targeting source Schechter function parameters and the survey parameters (angular/spectral resolution and sensitivity). Here we provide the implementation details of the conversion from the observed quantities, which come with different units in the literature, to the final source luminosity, in $L_\odot$ or erg s$^{-1}$. 

\begin{itemize}
\item Comoving voxel size $V_{\rm vox}$\\
Consider that the targeting spectral line has the rest frequency $\nu_{\rm rest}$ at redshift $z$. The survey has the angular pixel size $\Omega_{\rm pix}$ (we use the beam size instead if the survey does not specify their pixelization) and the spectral resolution $R=\nu_{\rm obs}/\delta\nu_{\rm obs}$, where $\nu_{\rm obs}=(1+z)\nu_{\rm rest}$ is the observed frequency. Then, the comoving voxel size is 
\begin{equation}
V_{\rm vox}= \Omega_{\rm pix}\left [ D_A^{CM}(z) \right ]^2\frac{c\,(1+z)}{H(z)\,R}
\end{equation}
where $c$ is the speed of light, $H(z)$ is the Hubble parameter, and $D_A^{CM}(z)$ is the comoving angular diameter distance, which equals to the comoving distance in the flat ($\Omega_k=0$) universe.

\item Deriving $L_{\rm SN}$ from the Schechter parameters\\
With the comoving voxel size and the luminosity function, we can calculate the $\sigma_{\rm SN}(\ell)$ following Eq.~\ref{E:sig_SN},
\begin{equation}
\sigma^2_{\rm SN}(\ell)=V_{\rm vox}\, \phi_\ast \int_0^\ell d\ell'\, \ell'^{\alpha+2}\,e^{-\ell'},
\end{equation}
and we find out $L_{\rm SN}$ numerically with the definition $\sigma_{\rm SN}(L_{\rm SN})=L_{\rm SN}$. 
\item Deriving $\sigma_L$ from the experiment sensitivity\\
The conversion of the instrumental noise to $\sigma_L$ is derived by matching the rms of noise flux $F^n$ to the source emission line flux $F^s$. Below we will work with flux in defined by power per area (in the units of W m$^2$). The flux $F^s$ from a line luminosity $\ell$ source is given by
\begin{equation}\label{E:L2F}
F^s=\frac{\ell}{4\pi D_{L}^2(z)},
\end{equation}
where $D_L(z)$ is the luminosity distance. As for the noise, if it is quoted as the ``flux density'' $F^{n}_\nu$[$erg/s/cm^2/Hz$], the noise flux $F^n$ is given by 
\begin{equation}
F^n=F^n_\nu \,\delta \nu_{\rm obs}=F_\nu\, (\nu_{\rm obs}/R).
\end{equation}
The $\sigma_L$ is then defined by the $\ell$ scale where $F_s=F_n$, and thus 
\begin{equation}\label{E:F2sigL}
\sigma_L=4\pi \,D_L^2(z)\,F^n_\nu\, \nu_{\rm obs}/R.
\end{equation}
If the sensitivity is quoted in $m_{AB}$ instead, then the flux density $F_{\nu}^n$ is given by $F^n_\nu=3631\times 10^{-m_{AB}/2.5} [Jy]$. If this is the $5\sigma$ sensitivity, then we use $F^n_\nu/5$ in the $\sigma_L$ calculation in Eq.~\ref{E:F2sigL}.

If the noise level is quoted in intensity $I^{n}_{\nu}[Jy/sr]$, then the conversion to the noise flux density per voxel is $F^{n}_{\nu}=I^{n}_{\nu}\Omega_{\rm pix}$. Finally, when noise is in the units of brightness temperature $T$,  the intensity $I^{n}_{\nu}$ can be derived using $I^{n}_{\nu}=2\nu_{\rm obs}k_BT/c^2$, and then we can get $\sigma_L$ with the equations listed above. 

\item Velocity-integrated luminosity\\
\citet{2016MNRAS.461...93P} quote their CO luminosity function in the ``velocity-integrated luminosity'' $L^{V}$ (Jy km s$^{-1}$ Mpc$^2$), which is the ``luminosity density'' (in units proportional to W Hz$^{-1}$) per observed velocity. To convert it to the intrinsic luminosity unit [$L_\odot$], we use the formalism in \citet{2009ApJ...702.1321O} Appendix A:
\begin{equation}
\frac{L}{L_\odot}=1.040\times 10^{-3} \left ( \frac{\nu_{\rm obs}}{\rm GHz} \right )\left ( \frac{1+z}{4\pi} \right )\frac{L^{V}}{\rm Jy\ km\ s^{-1}\ Mpc^2}
\end{equation}

\item HI mass-to-light ratio\\
To convert the HI mass function to the luminosity function, we follow the equation in \citet{2011piim.book.....D} in the optically thin limit, 
\begin{equation}
M_{\rm HI}=4.945\times 10^7 M_\odot \left ( \frac{D_L}{Mpc} \right )^2\left ( \frac{F^s}{\rm Jy\ MHz} \right ).
\end{equation}
Combining with Eq.~\ref{E:L2F}, we obtain the mass-to-light ratio
\begin{equation}
\frac{M_{\rm HI}}{M_\odot}=1.56\times 10^{8}\frac{L_{\rm HI}}{L_\odot}.
\end{equation}

\item CHIME instrument noise \\
We calculate the CHIME instrument noise using the parameters in \citet{2010ApJ...721..164S}. The noise rms per voxel is (in the temperature unit)
\begin{equation}
\sigma_T=\frac{gT_{\rm sky}+T_a}{\sqrt{t_{\rm int}\,\Delta f}}
\end{equation}
where $g$ is the gain and $T_{\rm sky}$ and $T_{\rm a}$ are the sky and antenna temperature, respectively. $\Delta f$ is the bandwidth, and $t_{\rm int}$ is the integration time  per pixel:
\begin{equation}
t_{\rm int}=N_{\rm year}\,D_f\,\frac{1}{2\pi}\frac{\lambda_{\rm obs}}{W_{\rm cyl}}
\end{equation}
where $N_{\rm year}$ is the total integration time, $D_f$ is the duty factor, $\lambda_{\rm obs}$ is the observed wavelength (42 cm at $z=1$), and $W_{\rm cyl}$ is the width of the cylinder. We use the parameter values listed in \citet{2010ApJ...721..164S}: $N_{\rm year}=1.4$ yr, $D_f=0.5$, $W_{\rm cyl}=14.3$ m, which gives $t_{\rm int}=3.3\times 10^{-3}$ yr. Then, we take $T_{\rm sky}=50$ K, $T_{\rm a}=10$ K, $g=0.8$, $\Delta f=390$ kHz, and we get $\sigma_T=2.9\times 10^{-4}$ K.

\end{itemize}

\end{appendices}

\bibliography{reference}

\end{document}